\documentclass[epjST]{svjour}
\usepackage{amsmath}
\usepackage{amssymb}
\usepackage{graphics}
\usepackage{color}
\usepackage[colorlinks=true,linkcolor=blue,citecolor=blue, urlcolor=blue]{hyperref}   
\usepackage[toc,page]{appendix}

\newcommand{\ddx}{\text{d}^4 x}
\newcommand{\mn}{{\mu \nu}}
\newcommand{\ab}{{\alpha\beta}}
\newcommand{\ie}{i.e.~}
\newcommand{\eg}{e.g.~}

\newcommand{\sigs}{\Sigma}

\newcommand{\gB}{G} 
\newcommand{\gt}{\tilde{g}}
\newcommand{\pt}{\tilde{p}}

\newcommand{\gb}{G}

\newcommand{\at}{\tilde{a}}
\newcommand{\bt}{\tilde{b}}
\newcommand{\ct}{\tilde{c}}

\newcommand{\com}[1]{\textcolor{black}{#1}}

\usepackage[demo]{graphicx}
\usepackage{subcaption}

\begin{document}
\title{Hydrodynamic attractors in heavy ion collisions: a review}
\author{Alexander Soloviev 
}                     
%
%
\institute{Center for Nuclear Theory, Department of Physics and Astronomy, \\Stony Brook University, Stony Brook, New York 11794, USA}
\date{Received: date / Revised version: date}
%
\abstract{A review of the recent progress of relativistic hydrodynamic attractors is presented, with a focus on applications in heavy ion collisions and the quark gluon plasma. Pedagogical introductions to the effective descriptions relevant for attractors in high energy physics, namely hydrodynamics, holography and kinetic theory, are followed by highlights of some recent advances.
%
} 
\maketitle

\section{Introduction}
\label{intro}

Relativistic hydrodynamics \cite{Kovtun:2012rj,Romatschke:2017ejr,Shen:2020mgh} is an incredibly useful tool in effective theories, capturing the underlying symmetries of a system in the long wavelength limit. It has been successfully applied at a vast swathe of scales: from the smallest in understanding quantum chromodynamics (QCD) at high energies \cite{Heinz:2013th}, to condensed matter systems \cite{chaikin2000principles} and to the largest in cosmological applications \cite{shore,Weinberg:2008zzc}. 

The application of hydrodynamics is not always readily apparent. In fact, in the high energy context one of the most unexpected surprises to come from evaluating the data at the Relativistic Heavy Ion Collider (RHIC) and the Large Hadron Collider (LHC) was that the quark-gluon plasma (QGP), the extreme state of matter composed from liberated constituents of neutrons and protons following collisions of heavy ions, behaves like the ``most perfect fluid'' and as such is amenable to a description via relativistic hydrodynamics. 
Abundant evidence points to the QGP exhibiting collective flow, such as the observation of elliptic 
flow \cite{Teaney:2009qa,Snellings:2011sz}
and the particularly low value of the ratio of the shear viscosity to entropy density \cite{Romatschke:2007mq}, close to the value predicted from holography \cite{Policastro:2001yc,Kovtun:2003wp,Kovtun:2004de}.
For further mechanisms and features of the thermalization of the QGP, see \cite{Berges:2020fwq} and references therein.

As such, the hydrodynamic description of the QGP had great phenomenological success. Active research deals with the hydrodynamization, i.e. the approach of the post-collision state to a fluid description, see \cite{Florkowski:2017olj} for a review. 
It is worth reiterating how surprising this is since hydrodynamics is an effective theory based on the \com{standard} assumption that the system is near thermal equilibrium, \com{although what exactly sets the regime of applicability of hydrodynamics was not clear.}
In typical heavy ion collisions, the system is certainly not in or even near equilibrium, which naturally beckons the question: why is hydrodynamics so unreasonably effective?

Recently, there has been progress in understanding how systems with little hope of being in equilibrium become nonetheless well-approximated by hydrodynamics. One proposed solution is 
due to the emergence of attractor phenomena. In a word, the study of attractors stems from the observation that solutions of evolution equations with different initial conditions evolve to the same late time behavior, which is called the (hydrodynamic) attractor. Quite generally, nonequilibrium solutions of different microscopic theories decay to attractors, losing knowledge of their initial conditions well before reaching thermal equilibrium. Specifically, heavy ion collisions have a mix of hydrodynamic and non-hydrodynamic modes, the latter of which decay rapidly in the expanding medium, leaving behind just the long wavelength modes, well described by hydrodynamics.
For the heavy ion community, attractors came to the forefront with the remarkable discovery of the hydrodynamic attractor in an expanding boost-invariant  background by Heller and Spalinski 
\cite{Heller:2015dha},
building on the earlier results of Borel resummed hydrodynamic gradient expansion in holographic Bjorken flow \cite{Heller:2011ju,Heller:2013fn}.

Spurred on by the characterization of the hydrodynamic attractors, attention was drawn to attractors in different descriptions relevant for heavy ion collisions. 
For instance, similar attractors were found in kinetic theory and in holography, even in non-conformal and non-homogeneous settings \cite{Florkowski:2017jnz,Romatschke:2017acs}, which indicates that attractor solutions should exist for a wide range of theories independent of the symmetries of those theories. 
In hindsight, this should not be so surprising, since the hydrodynamic limit is found in so many diverse physical settings. For instance, hydrodynamic attractors have also been explored in holographic models of QGP \cite{Buchel:2016cbj} 
and in Gauss-Bonnet gravity \cite{Casalderrey-Solana:2017zyh}, which could have well been anticipated due to the fluid-gravity correspondence \cite{Rangamani:2009xk}. 

 A key feature of the attractor is that when it is present, the system ``forgets'' about its initial conditions. Therefore, attractors can be used phenomenologically to distinguish between observables which are dependent on initial conditions and those which are not. This can lead to the identification of universal behavior in heavy ion collisions. 
Recently, the potential of hydrodynamic attractors as a tool for phenomenological studies has begun to be appreciated.  For instance, they have been used to relate initial state energies to the produced particle multiplicity \cite{Giacalone:2019ldn,Jankowski:2020itt}. Moreover, recent studies examined the hydrodynamic attractor at finite chemical potential \cite{Dore:2020fiq,Dore:2020jye}, tracing the hydrodynamic evolution as trajectories through the QCD phase diagram.
One of the more exciting ideas relevant to heavy ion physics to come to fruition at least in part due to attractors is the Kurkela-Mazeliauskas-Paquet-Schlichtling-Teaney (K{\o}MP{\o}ST) framework \cite{Kurkela:2018wud,Kurkela:2018vqr}, essentially linking the early time dynamics, captured by a kinetic description, to the late time hydrodynamic behavior.
Another practical application of hydrodynamic attractors is as a viable arena to study resurgence  \cite{Basar:2015ava,Aniceto:2018bis}. A comprehensive discussion of resurgence is beyond the scope of the present review.

Of course, it would be remiss to not mention that attractors have a rich, storied history in dynamical systems. There is considerable value in trying to understand the attractor studied in heavy ion collisions in the more mathematical language of dynamical systems. Progress in this direction has already been made, see \cite{Behtash:2017wqg,Behtash:2019txb,Behtash:2019qtk}.

A related story to the hydrodynamic attractor is the study of non-thermal fixed points \cite{Berges:2013eia,Berges:2013fga,Berges:2013lsa} in heavy ion physics (as well as in a variety of other contexts \cite{Berges:2014bba}), which are essentially an observation that at early enough times, the distribution function exhibits a special scaling behavior. This early time regime is currently the subject of numerical study, such as in a simplified model of the glasma in $2+1$ dimensions \cite{Boguslavski:2019fsb}. This should be considered as a distinct feature of the weakly coupled degrees of freedom at early times in the QGP, different from the hydrodynamic attractors that we focus on here. 

Since hydrodynamic attractors have been found in a wide range of (effective) theories pertinent to heavy ion collisions, many works have focussed on comparing features of attractor behavior in relativistic hydrodynamics, kinetic theory and holography \cite{Romatschke:2017acs,Kurkela:2019set}. These theories vary not only in their frameworks, but also in their description of weak and strong coupling regimes. The benefits of this approach are clear, giving ample evidence of the importance of the attractors. 
This review will present in pedagogical detail the attractor story for the trio of theories mentioned previously, first by describing an illustrative example, followed by reviewing current progress in the field. 

As such, the organization of the paper is as follows. In Sec.~\ref{sec:hydro}, we will discuss the prototypical example of the hydrodynamic attractor in the simplest setting of $(0+1)$-dimensional Bjorken expansion, before reviewing further progress.  Next, in Sec.~\ref{sec:holo}, we will present the attractor as it pertains to holographic systems. Then, we will turn our attention to attractors in kinetic theory in Sec.~\ref{sec:kt}, especially focussing on the relaxed time approximation. Finally, in Sec.~\ref{sec:pheno}, we turn to the recent applications of attractors to heavy ion collisions. A table of useful abbreviations is included in Appendix \ref{app:abbrev}.

\section{Hydrodynamic attractors}\label{sec:hydro}

In this section, we will review the advances in studying the attractor in expanding relativistic hydrodynamics. As is usually the case, it is best to start simple and then discuss more complicated examples. We will refresh the reader with a discussion of relativistic hydrodynamics and  various models of curing the acausality, before reaching the simplest and quintessential example of an attractor in $0+1d$ M{\"u}ller-Israel-Stewart (MIS) Bjorken flow. We will finish by going into some detail of some advances in the field.

\subsection{Primer on relativistic hydrodynamics}

Hydrodynamics represents the course-grained description of microscopic dynamics. The macroscopic objects of interest are the energy momentum tensor, $T^\mn$, and the conserved currents, $J^\mu$. The dynamical equations are generated by considering the conservation of the energy momentum tensor,
\begin{align}\label{cons}
\nabla_\mu T^\mn=0,
\end{align}
as well as the conservation of the currents, schematically $\nabla_\mu J^\mu=0$.

Generally, the fluid energy momentum tensor can be decomposed into ideal and dissipative contributions
\begin{align}\label{fluid-emt}
T^\mn= T^\mn_{\text{ideal}}+\Pi^\mn.
\end{align}
The ideal energy-momentum tensor is
\begin{align}
T^\mn_{\text{ideal}}=\varepsilon u^\mu u^\nu+P \Delta^\mn,
\end{align}
where $\varepsilon$ is the energy density, $P$ is the pressure and $\Delta^\mn =u^\mu u^\nu +g^\mn$ is the spatial projection operator. Note that the four-velocity, $u^\mu$, is timelike normalized, $u^\mu u_\mu=-1$. The dissipative tensor, $\Pi^\mn$, generically contains higher derivative corrections of $u^\mu$ and the temperature, $T$. This organization is known as the \textit{gradient expansion}. 

We will ultimately need to specify a reference frame for the fluid. Here, we choose to work in the \textit{Landau frame}, where the energy density is the eigenvalue associated with $u^\mu$ via $T^\mn u_\nu=-\varepsilon u^\mu$. This leads to the requirement that the dissipation tensor is orthogonal to the flow, \ie{}
\begin{align}
\Pi^\mn u_\mu=0.
\end{align}

At this point, the standard approach would be to specify the dissipation tensor order by order in a gradient expansion with arbitrary transport coefficients. These transport coefficients are part of the required input for a complete hydrodynamic description, which can be determined from the underlying microscopic theory. To first order in derivatives, we have two distinct tensor structures, which we write as follows
\begin{align}\label{diss1}
\Pi^\mn =-\eta\sigma^\mn -\zeta \Delta^\mn \nabla_\alpha u^\alpha,
\end{align}
where the two transport coefficients are the shear viscosity, $\eta$, and the bulk viscosity, $\zeta$. Higher orders lead to more possible tensor structures with additional transport coefficients. The shear tensor is transverse, 
\begin{align}
\sigma^\mn u_\mu=0,	
\end{align}
traceless, \begin{align}
 g_\mn \sigma^\mn=0,	
 \end{align}
 and is explicitly given by
\begin{align}
\sigma^\mn=\Delta^{\mu\alpha}\Delta^{\nu\beta}(\nabla_\alpha u_\beta+\nabla_\beta u_\alpha)-\frac{2}{3}\Delta^\mn\nabla_\alpha u^\alpha.
\end{align}

The causal hydrodynamic evolution of a system can be described via \eqref{cons}
. There is one complication: the four equations in \eqref{cons} contain five variables: the energy density, $\varepsilon$, the pressure, $P$, and the four velocity, $u^\mu$. As mentioned before, the last needs to satisfy the timelike normalization condition $u^\mu u_\mu=-1,$ meaning that $u^\mu$ has three independent components. Thus, we are shy one equation. 

To fill this gap, we will need to specify an equation of state, relating the pressure to the energy density, $P=P(\varepsilon)$. In general, the equation of state depends on the system that one is interested in modelling. One of the simplest choices (which we will largely adopt here) is the conformal equation of state $\varepsilon=3P$, which means that the energy momentum tensor is traceless, \begin{align}
g_\mn T^{\mu\nu}_{\rm }=0. 	
 \end{align}
 In the conformal case, the energy density can only depend on temperature, which means that $\varepsilon \sim T^4.$

\subsubsection{Truncation schemes}

The issue with the choice of \eqref{diss1} for the dissipation tensor is that resulting equations become acausal \cite{Hiscock:1985zz,Kostadt:2000ty}. A straightforward way to see this, e.g. \cite{Romatschke:2017ejr}, is by considering metric perturbations of the form, $g_\mn=\eta_\mn+h_\mn$, and seeing how the stress tensor \eqref{fluid-emt} changes due to the metric perturbation in linear response. In essence, we are computing the retarded correlator, $G^{\mn,\ab}_R$, via
\begin{align}
T^\mn=T_0^\mn -\frac{1}{2}G^{\mn,\ab}_R h_\ab+\ldots
\end{align}
where the dots obscure terms higher order in perturbations. Focussing simply on the shear channel, we see that the retarded correlator is \cite{Kovtun:2012rj}
\begin{align}
G^{0x,0x}_R=\varepsilon+\frac{\eta k^2}{i\omega -\frac{\eta}{\varepsilon+P}k^2}.
\end{align}
From the denominator, we identify the diffusive pole, $\omega=-i \frac{\eta}{\varepsilon+P}k^2$. The group velocity, $\left\vert{d\omega}/{dk}\right\vert$, is given by 
\begin{align}
\left\vert\frac{d\omega}{dk}\right\vert=2\frac{\eta}{\varepsilon+P}k,
\end{align}
which grows without bound as $k$ gets large.

There are a few ways to cure such acausality. A simple and popular approach is to implement the MIS equations \cite{Muller:1967zza,Israel:1976tn,Israel:1979wp} (for more recent developments, see  Baier-Romatschke-Son-Starinets-Stephanov (BRSSS) \cite{Baier:2007ix}), where the dissipation tensor satisfies a relaxation equation. Here, as we will be interested in the conformal case of non-zero shear viscosity, but vanishing bulk viscosity (as well as higher order transport coefficients), the MIS equation in this case is then
\begin{align}\label{mis}
\left(\tau_\pi u^\alpha \nabla_\alpha+1\right)\Pi^\mn=-\eta \sigma^\mn,
\end{align}
where $\tau_\pi$ is the relaxation time. The previous case \eqref{diss1} corresponds to vanishing relaxation time $\tau_\pi$ (as well as $\zeta=0$). 

How does this cure the acausality? For small only time-dependent perturbations, we see that 
\begin{align}
\delta \Pi^\mn=\frac{\eta}{i\tau_\pi \omega-1}\delta \sigma^\mn,
\end{align}
which can be compared directly to \eqref{diss1} in the limit that $\omega,\zeta\rightarrow0$.
This simply amounts to replacing $\eta\rightarrow \frac{\eta}{i\tau_\pi \omega-1}$. Revisiting the computation for the retarded correlator, we see that the diffusive mode is found when 
\begin{align}
i\omega(1-i \tau_\pi\omega)-\frac{\eta}{\varepsilon+P}k^2=0.
\end{align}
Solving for $\omega$ and computing the group velocity, we find one of the modes has dispersion
\begin{align}
\left\vert\frac{d\omega}{dk}\right\vert=\frac{2\frac{\eta}{\varepsilon+P} k}{\left\vert\sqrt{ \frac{4\eta\tau_\pi}{\varepsilon+P}k^2 -1}\right\vert}.
\end{align}
The group velocity now saturates to a maximum value, namely
\begin{align}\label{max-speed}
\lim_{k\rightarrow\infty}\left\vert\frac{d\omega}{dk}\right\vert=\sqrt{\frac{\eta}{\tau_\pi(\varepsilon+P)}}
\end{align}
and as such, there is no issue with causality, so long as the right hand side of \eqref{max-speed} is smaller than 1. 

A more recent improvement to the MIS equations is known as Denicol-Niemi-Molnar-Rischke (DNMR) effective theory \cite{Denicol:2012cn}, which is a second order hydrodynamic theory using Grad's 14-moment approximation for the single particle distribution function. To first order in gradients, DNMR is similar to MIS \eqref{mis}, in that (taking $u^\mu=(1,0,0,0)$) we see that the general form the equations take to this order is \cite{Strickland:2017kux} 
\begin{align}\label{DNMR}
\partial_\tau{\Pi}=\frac{4\eta}{3\tau\tau_\pi}-\beta_{\pi\pi}\frac{\Pi}{\tau}-\frac{\Pi}{\tau_\pi},
\end{align}
where for DNMR we have $\beta_{\pi\pi}=38/21$ and $\tau_\pi=\tau_{eq}$ \cite{Denicol:2012cn}, while for MIS this is $\beta_{\pi\pi}=4/3$ and $\tau_\pi=6\tau_{\rm eq}/5$ \cite{Muronga:2003ta}.

Another truncation scheme goes under the name of anisotropic hydrodynamics (aHydro) \cite{Strickland:2014pga,McNelis:2018jho,Almaalol:2018ynx,Alalawi:2020zbx}. Essentially, the idea is to formulate hydrodynamics in momentum space with a view towards taking into account the large anisotropies found in the QGP. One takes a distribution function with an explicit anisotropy parameter, $\xi$ (see Sec.~\ref{sec:kt} for a primer on kinetic theory and \eqref{aniso-dist} for the form of the anisotropic distribution function). As outlined in Sec.~\ref{sec:kt}, the Landau matching condition determines the temperature. The evolution of $\xi$ is determined by considering a higher moment of the Boltzmann equation.
In this way, aHydro resums an infinite number of terms in an expansion in inverse Reynolds number \cite{Strickland:2017kux}
\begin{align}
R^{-1}=\frac{\sqrt{\Pi^\mn \Pi_\mn}}{P_0}\propto \bar{\Pi}(\xi),
\end{align}
where $\bar{\Pi}=\Pi/\varepsilon.$ A comparison to the other truncation schemes can be made by rewriting the evolution equation for $\xi$ as a function of $\Pi$ and expanding for small anisotropy. This results in agreement with \eqref{DNMR} to linear order in $\Pi$. However, note that at higher order $\Pi^2$, the term $\lambda_1 \Pi^2 / (2\tau_\pi \eta^2)$ has a numerical disagreement of the transport coefficient, $\lambda_1^{\rm aHydro}=\eta\tau_\pi/7$, with $\lambda_1^{\rm DNMR}=5\lambda_1^{\rm aHydro}$ \cite{Romatschke:2011qp,Strickland:2017kux}. The differences between the two approaches arises from the fact that aHydro includes higher order terms compared to DNMR at the same level of truncation.  

Which scheme to use? In terms of conceptual simplicity, the MIS framework tends to get more mileage since it is rather straightforward to include additional higher order transport coefficients. 
MIS and DNMR differ mainly in their far-from-equilibrium transient effects, as they have the same late time Navier-Stokes limit \cite{Dore:2020jye}.
It was found in \cite{Strickland:2017kux} that DNMR is more accurate than MIS when compared to the exact solution in the relaxed time approximation (RTA), as it captures systematically second order hydrodynamic contributions, while aHydro approximates the exact attractor the best. Likewise, in the context of Gubser flow where the exact attractor is known,
aHydro did a better job in capturing the attractor compared to MIS and DNMR \cite{Heinz:2015gka,Behtash:2017wqg}.

As an aside, it is worthwhile to underline that the MIS equations represent a particularly easy model to work with. However, if one were inclined to include non-hydrodynamic modes to their model building, such as the oscillatory behavior of $N=4$ super Yang-Mills theory via a hydrodynamic model, the Heller-Janik-Spalinski-Witaszczyk (HJSW) model could be considered a natural generalization to the MIS equations \cite{Heller:2014wfa,Aniceto:2015mto,Heller:2020anv}. In this case, the dissipation tensor, $\Pi^\mn_{\rm total}=\Pi^\mn_{\rm MIS}+ \Pi^\mn_{\rm HSJW}$, can be decomposed into a piece evolving with the MIS equations \eqref{mis} and another piece satisfying\footnote{Note that for the purposes of modelling, one can also study the case where $\Pi^\mn_{\rm MIS}=0.$}
\begin{align}\label{hsjw}
\left((\frac{1}{T}u^\alpha \nabla_\alpha)^2
+2\Omega_I u^\alpha\nabla_\alpha +\vert\Omega\vert^2\right)\Pi^\mn_{\rm HSJW}=-\eta \vert \Omega\vert^2 \sigma^\mn-\frac{c_\sigma}{T} u^\mu \nabla_\mu(\eta \sigma^\mn)+\ldots
\end{align}
where $\Omega=\Omega_R+i\Omega_I$ is the complex frequency, meant to mimic the quasinormal mode frequency. The values of the real and imaginary part can be fixed by comparing to $N=4$ Super Yang-Mills (SYM) theories \cite{Nunez:2003eq,Aniceto:2015mto}. The presence of these non-hydrodynamic modes is guaranteed by the appearance of the second derivative in the above. In principle, one can further generalize by considering higher derivatives on the left hand side of \eqref{hsjw}, although how to specify the extra initial conditions is not clear.

\subsection{$0+1d$ Bjorken flow}\label{bjork-example}

Now let's get our hands dirty with an example computation. We will review the hydrodynamic attractor in Bjorken flow \cite{Heller:2015dha}, a $(0+1)$-dimensional
expanding transversally homogeneous system with longitudinal boost-invariance.

Bjorken flow \cite{Bjorken:1982qr} describes a simple model for heavy ion collisions, where the collision
axis is along the z-axis and the system is otherwise homogeneous. Essentially, the nuclei are
modelled as infinite flat sheets in the transverse ($x, y$) directions. Furthermore, the matter
produced in the forward light cone is boost invariant. The background metric is given in Milne
coordinates 
\begin{align}\label{bjorken-metric}
g_\mn=\text{diag}(-1,1,1,\tau^2).
\end{align}
We can relate Milne coordinates $(\tau,x,y,\xi)$ to Minkowski coordinates $(t,x,y,z)$ by observing that the proper time and the spacetime rapidity are
\begin{align}
\tau&=\sqrt{t^2-z^2},\\
  \xi&=\tanh^{-1} (z/t),
 \end{align}
 respectively. Note that the metric \eqref{bjorken-metric} describes a space which is Ricci flat. For future convenience, we also note that the nonzero Christoffel symbols are 
\begin{align}
\Gamma^{\tau}_{\phantom{\tau}\xi \xi}=\tau, \quad \Gamma^{\xi}_{\phantom{\xi}\xi\tau}=\tau^{-1}.
\end{align}

We will assume all variables depend only on the proper time, $\tau$, which means that \eqref{cons} and \eqref{mis} take a particularly simple form:
\begin{align}\label{ward0}
0&=\tau \dot\varepsilon+\frac{4}{3}\varepsilon-\phi,\\
0&=\tau_\pi \dot \phi+\frac{4 \tau_\pi \phi}{3\tau}+\phi-\frac{4\eta}{3\tau},\label{mis0}
\end{align}
where $\phi\equiv-\Pi^\xi_\xi$. Thus, the conservation equation, \eqref{ward0}, and the MIS equation, \eqref{mis0}, represent the set of equations describing the relativistic evolution of a viscous fluid in a Bjorken expanding background. We proceed by expressing the shear viscosity and the relaxation time in terms of the energy density:
\begin{align}
\eta = C_\eta \frac{\varepsilon+P}{\varepsilon^{1/4}}, \quad  \quad \tau_\Pi= C_{\tau_\Pi} \varepsilon^{-1/4},
\end{align}
where $C_\eta$ and $C_{\tau_\Pi}$ are dimensionless constants. \com{As we will see shortly, the relaxation time is the hydrodynamic transport coefficient which indicates the decay time of the non-hydrodynamic mode.}
 Taking insight from holography \cite{Baier:2007ix}, we will take these constants to be
\begin{align}\label{ceta}
C_\eta=\frac{1}{4\pi}, \quad C_{\tau_\Pi}=\frac{2-\ln 2}{2\pi}.
\end{align} 
We can then eliminate $\phi$ to arrive to arrive at an expression solely for the energy density
\begin{align}
9 C_{\tau_\Pi} \tau ^2\ddot{\varepsilon}+(33 C_{\tau_\Pi} \tau +9 \tau ^2 \varepsilon^{1/4}  )\dot\varepsilon+16 (C_{\tau_\Pi}-C_{\eta}) \varepsilon
+12 \tau\varepsilon^{5/4}=0.
\end{align}
We can recover the expressions found in \cite{Heller:2015dha} by remembering that in a conformal theory the energy density is related to the temperature via $\varepsilon \sim T^4$.

\begin{figure}\centering
\includegraphics[width=0.75\linewidth]{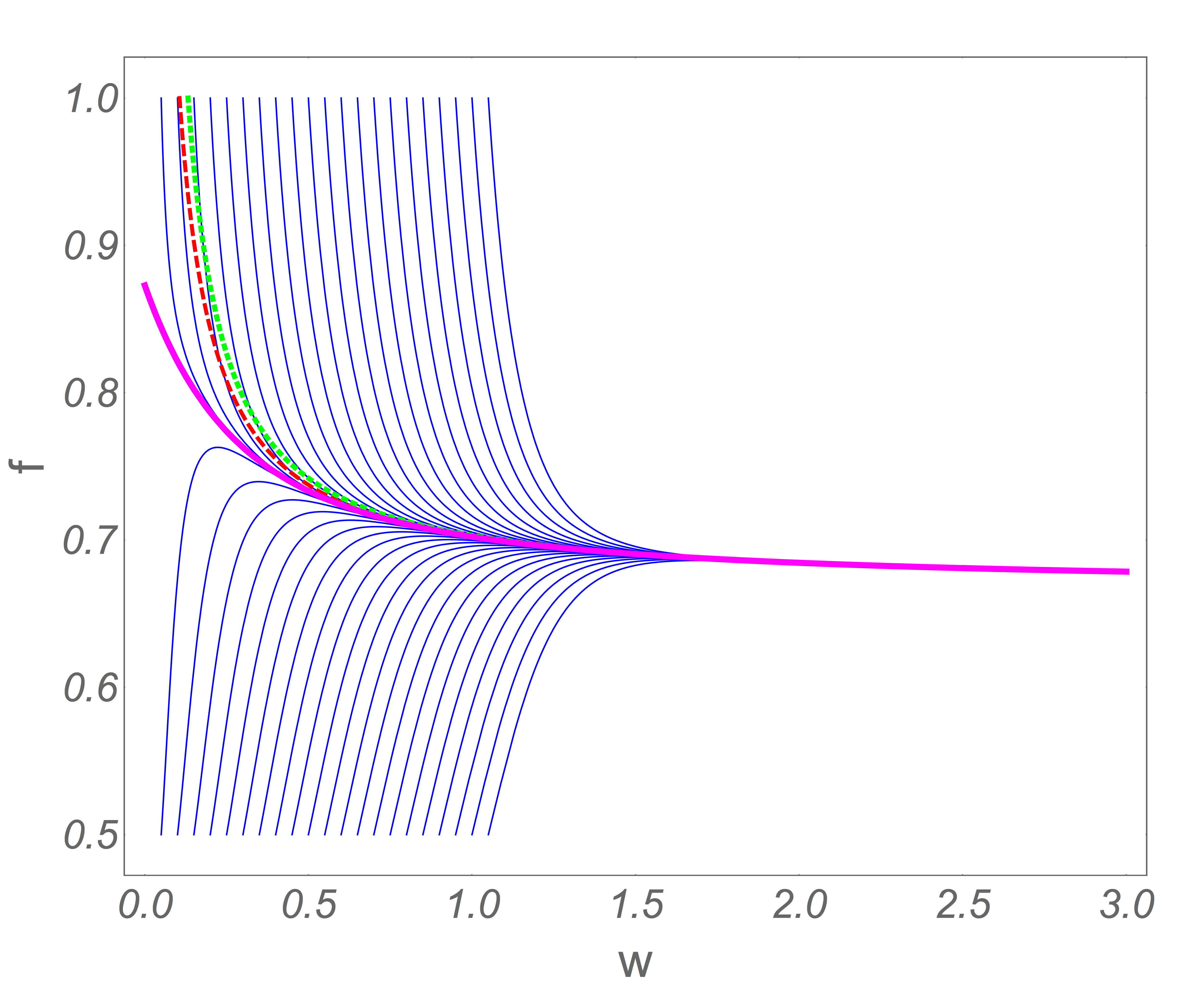}
\caption{Hydrodynamic attractor for $0+1d$ Bjorken flow. Solutions of \eqref{eq-f0} with various initial conditions are shown in blue, while the attractor is shown in magenta. The red and green dashed lines correspond to first and second order hydrodynamics. Figure from \cite{Heller:2015dha}.}
\label{fig:hydro}
\end{figure}

Following \cite{Heller:2011ju,Heller:2015dha}, we can now introduce dimensionless quantities 
\begin{align}\label{fdef}
\omega=\tau \varepsilon^{1/4},  \quad f=\tau \partial_\tau \log{\omega},
\end{align} 
which leads to the following equation for $f=f(\omega)$
\begin{align}\label{eq-f0}
C_{\tau_\Pi}  \omega  f  f^\prime (\omega)
+4 C_{\tau_\Pi} f^2 
 +\left(  \omega -\frac{16 C_{\tau_\Pi}}{3} \right) f
-\frac{4 C_\eta  }{9}
+\frac{16 C_{\tau_\Pi} }{9}-\frac{2  \omega }{3}=0.
\end{align}
We can now numerically solve the above equation for a variety of initial conditions, which we plot in Fig.~\ref{fig:hydro}. Irrespective of the choice of initial conditions, the large $\omega$ behavior converges to a single curve, which is known as the hydrodynamic attractor. In this concrete sense, the attractor is said to ``forget'' about its initial conditions. The approach to the attractor can be found if we solve the linearized form of \eqref{eq-f0} for $\delta f$. We find that
\begin{align}\label{lin-f}
\delta f =e^{-\frac{3 \omega }{2 C_{\tau_\Pi}}} \omega^{C_\eta/C_{\tau_\Pi}} +\mathcal{O}(\omega^{-1}).
\end{align}
From here it clear to see that the perturbation decays exponentially to the attractor. 

Note that another often used dimensionless quantity (see e.g. \cite{Florkowski:2017olj}), known as the pressure anisotropy, is defined as 
\begin{align}\label{anisotropy-parameter}
\mathcal{A}\equiv\frac{\mathcal{P}_T-\mathcal{P}_L}{\mathcal{E}/3},
\end{align}
which is normalized with respect to the equilibrium pressure, $P=\mathcal{E}/3$. The two dimensionless quantities are related simply via
\begin{align}
\mathcal{A}=18\left(f(\omega)-\frac{2}{3}\right).
\end{align}

\begin{figure}\centering
\includegraphics[width=0.75\linewidth]{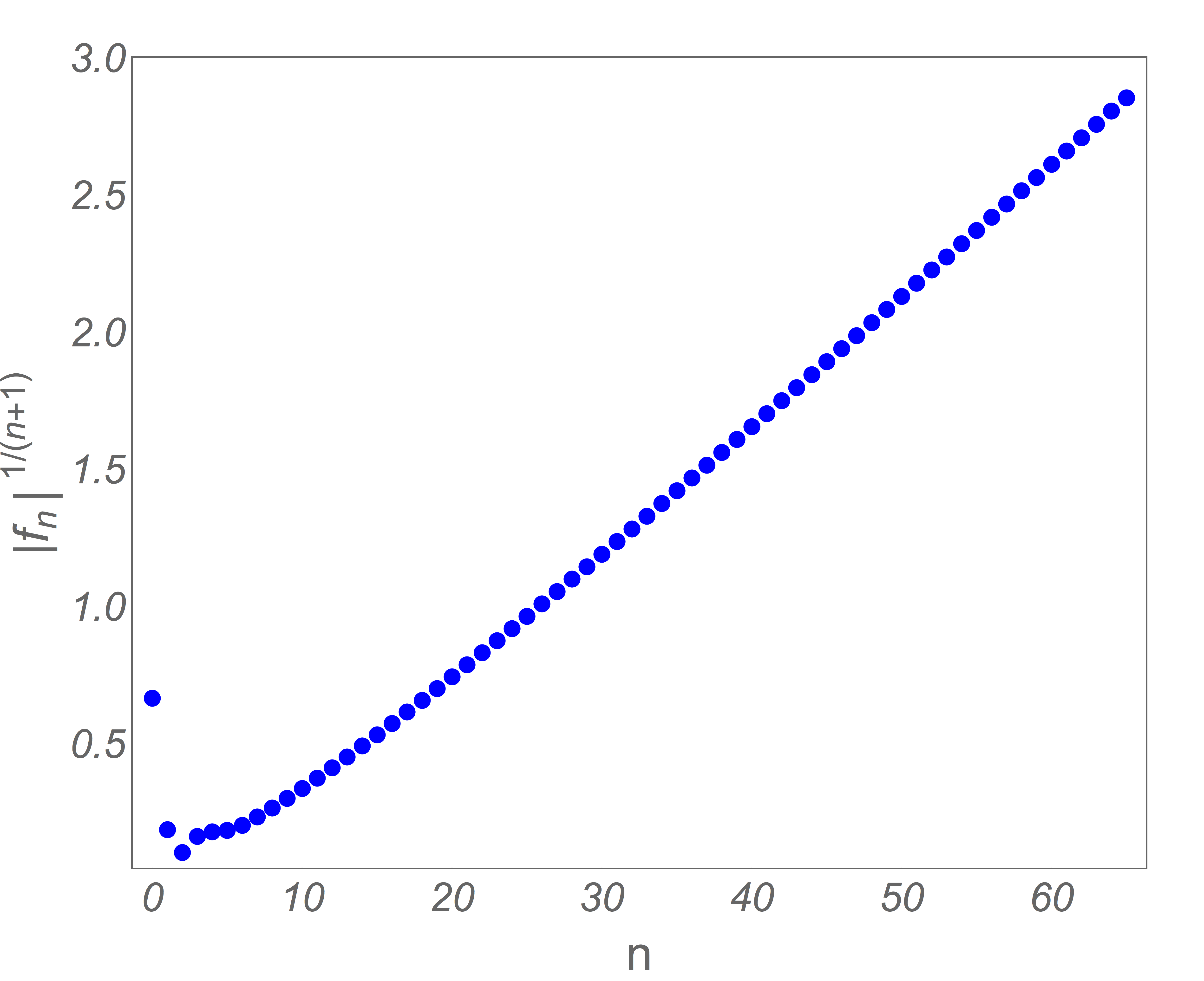}
\caption{ Coefficients in the expansion of $f$ \eqref{naive-exp} grow like $n!$ Figure from \cite{Heller:2015dha}.}
\label{fig:factorial}
\end{figure}

We can now examine the hydrodynamic gradient expansion as an asymptotic expansion in $\omega$ of \eqref{eq-f0}, given by
\begin{align}\label{naive-exp}
f&=\sum_{n=0}^\infty f_n\omega^{-n},\\
&=\frac{2}{3}+\frac{4C_\eta}{9 \omega}+\frac{8C_\eta C_{\tau_\Pi}}{27\omega^2}+\mathcal{O}\left(\omega^{-3}\right).
\end{align}
It is straightforward to see that the resulting series has terms growing factorially \cite{Heller:2013fn}, indicating that the series is indeed divergent, see Fig.~\ref{fig:factorial}.

It is worthwhile to point out that the choice of variables $T(\tau)\leftrightarrow f(\omega)$ has nothing to do with the appearance of the attractor, which is inherent to the system. For the case considered above, the attractor can alternatively be captured in a study of the phase space \cite{Heller:2020anv}, by studying for instance the evolution of $(\tau_0 T(\tau),\tau_0^2\dot{T}(\tau))$ as a function of $\tau$. As food for thought, the evolution of a set of initial conditions in phase space can be viewed as dynamical dimensional reduction of the phase space regions. This might be a helpful way to think about the attractor in more realistic scenarios, e.g. in the case of relaxed symmetries.

Another method to characterize the attractor is known as the slow-roll approximation, originally studied in the cosmological setting \cite{Liddle:1994dx}. Formally, it amounts to expanding \eqref{eq-f0} in derivatives, by adding a small parameter, $\delta$, to the $f f^\prime$ term \cite{Strickland:2017kux}. One then expands \eqref{eq-f0} in powers of $\delta$
\begin{align}
f=\sum_{n=0}^\infty f_n \delta^n .
\end{align}
This is a useful computational technique, as at a given order, it contains derivative terms to all orders, although it should be noted that the slow-roll approximation diverges \cite{Denicol:2017lxn}.

The Bjorken case is one of the more simple examples due to the assumed symmetries and low dimensionality. The discussion presented lends itself naturally to generalizations, which we will review here after a brief aside.

\subsubsection{Quick introduction to trans-series}

Trans-series are an interesting mathematical development with real physics application. In some sense, they represent a generalization to Taylor series, taking into account not just the polynomial powers, but also the non-perturbative structure, such as exponentials and logarithms. A related topic, resurgence, is the curing of a perturbatively divergent series with non-perturbative terms \cite{Dorigoni:2014hea}. Resurgence can be used to deduce nonperturbative effects via analytic continuation from a perturbative series. For an introduction to trans-series with a more mathematical flavor, see \cite{transbegin}, and for more on resurgence, see for example \cite{Aniceto:2018bis}. 

Here, we will follow the discussion of trans-series relevant to hydrodynamics \cite{Basar:2015ava}, noting that the transseries of $N=4$ SYM in Bjorken flow have been explored in \cite{Aniceto:2018uik}.

As we see in \eqref{lin-f}, there is the non-hydrodynamic mode, exponentially supressed at late times, analogous to quasinormal modes in gravitational theory \cite{Aniceto:2015mto} and to instantons in quantum field theories \cite{Schafer:1996wv}
. It becomes clear that instead of the naive expansion \eqref{naive-exp}, one should seek a trans-series ansatz of the form 
\begin{align}\label{ftrans}
f(\omega)\sim \sum_{n=0}^\infty f^{(n)}(\omega)\sigma^n \zeta(\omega)^n,
\end{align}
where $\sigma$ is the trans-series expansion parameter, which counts the ``non-perturbative'' order of the trans-series and is generally complex and 
\begin{align}
\zeta(\omega)\equiv\omega^\beta e^{-S\omega},
\end{align}
is the so-called ``instanton fugacity''. 

Plugging this ansatz into \eqref{eq-f0} leads to equations at each order in $\sigma.$ For $n=0,$ we have the previous case discussed, namely the naive Taylor series. There is only freedom to fix one parameter, $f^{(1)}_0=1$, and all coefficients are then determined by their recursion relation, which is a generic feature of trans-series \cite{transcostin}.
The resulting non-hydrodynamic terms in the trans-series correspond to exponentially damped modes times a gradient expansion.
It is also worthwhile to point out that the overall constant is known as the Stokes constant, which when fixed resolves the ambiguity in the choice of contour when doing the inverse Borel transform \cite{Casalderrey-Solana:2017zyh} (note that in the example we consider in Sec.~\ref{sec:hybrid}, we would have two Stokes coefficients, while they are infinite for holography).

Do trans-series tell us anything about early times, i.e. when $\omega$ is small? The exponentially suppressed terms at asymptotically late times can become large compared to the powers of $\omega$ for early times, with such terms jockeying for position in size. Trans-asymptotic rearrangement rewrites the trans-series to better study the early time behavior \cite{transmatch}. For the example considered in \cite{Basar:2015ava}, 
we should rewrite the expansion \eqref{ftrans} as
\begin{align}
f(\omega)\sim\sum_{k=0}^\infty F_k (\sigma \zeta)\omega^{-k},
\end{align}
where $ F_k (\sigma \zeta)\equiv \sum_{k=0}^\infty f_k^{(n)}(\omega)\sigma^n \zeta^n.$ Inserting the reordered ansatz into \eqref{eq-f0}, one can solve order by order, finding that for $k>1$ each equation contains just a linear term in $F_k.$ The imaginary part of the parameter $\sigma$ is fixed in terms of the  Stokes parameter by the requirement that $f(\omega)$ is real. The real part of $\sigma$ is arbitrary \cite{Aniceto:2013fka}, thus parameterizing the different possible initial conditions at early times.


\subsection{Gubser flow}

A straightforward generalization of Bjorken flow was introduced by Gubser in \cite{Gubser:2010ze}, which has been studied at length in various settings. 
Like Bjorken flow, Gubser flow is boost-invariant with the additional aspect that there is azimuthally symmetric radial expansion. \com{In fact, a radially expanding Bjorken flow can be mapped to Gubser flow via an appropriate Weyl rescaling. It can be thought of as a comoving flow in conformal theories.}
The Gubser flow attractor has been widely studied as it is closer to the set up relevant in heavy ion collisions
\cite{Behtash:2017wqg,Bemfica:2017wps,Denicol:2018pak,Chattopadhyay:2018apf,Behtash:2019qtk,Dash:2020zqx}.

In terms of the de Sitter coordinates $(\rho, \theta, \phi, \eta)$, the Gubser flow metric is
\begin{align}\label{gubser}
g_\mn=\text{diag}(-1,\cosh^2\rho,\cosh^2 \rho \sin^2 \theta, 1),
\end{align}
which relates to the Milne coordinates via
\begin{align}
\sinh \rho&=-\frac{1-q^2\tau^2+q^2 r^2}{2q \tau},\\
\tan \theta &=\frac{2qr}{1+q^2\tau^2-q^2 r^2},
\end{align}
where $q^{-1}$ is the characteristic length scale of the transverse direction of the system. The time variable is $-\infty<\rho<\infty$. That the system is expanding is straightforward to see by computing
\begin{align}
\nabla_\mu u^\mu=2\tanh \rho.	
\end{align}

One then finds that the conservation of the energy momentum tensor, $\nabla_\mu T^\mn=0,$ satisfies a single equation, namely \cite{Marrochio:2013wla,Behtash:2017wqg}
\begin{align}\label{gubser-cons}
0=\frac{1}{T}\partial_\rho T+\frac{2}{3}\tanh{\rho}-\frac{1}{3}\pi \tanh\rho,
\end{align}
where the normalized shear stress is $\pi\equiv-\pi^{\eta\eta}/(\varepsilon+P)$ (note that this parameterization is different from $\phi$ introduced below \eqref{mis0}). 
As mentioned above, the evolution of $\pi$ can be specifed in a number of ways, the simplest of which is arguably MIS \eqref{mis}, which in Gubser coordinates reads \cite{Behtash:2017wqg} 
\begin{align}
0=\partial_\rho \pi + \frac{\pi}{\tau_\pi} +\left(\frac{4}{3} \pi^2 -\frac{4}{3}\frac{\eta}{s T \tau_\pi}\right) \tanh{\rho} .
\end{align} 
Using conformal symmetry, we can write $\tau_\pi=c/T$ with $c=5\eta/s$. 
As in the Bjorken case, the two equations can be combined into one equation by solving for $\pi$ in \eqref{gubser-cons}. Similar to the discussion in the Sec.~\ref{bjork-example}, a compact way to write the above equations is by introducing the dimensionless time $\omega=\tanh\rho/T$ and the anisotropy parameter \cite{Behtash:2017wqg}
\begin{align}
\mathcal{A}=\frac{1}{\tanh{\rho}}\partial_\rho \log T,
\end{align}
which is related via the conservation equations to the normalized shear stress as ${\pi}=3\mathcal{A}+2$.
We thus have
\begin{align}
	3\omega(\coth^2\rho-1-\mathcal{A}(\omega)) \mathcal{A} ^\prime(\omega)+\frac{4}{3}(3\mathcal{A}(\omega)+2)^2+\frac{3 \mathcal{A}(\omega)+2}{c\omega}-\frac{4}{15}=0
	\end{align}
As a useful aside, note that DNMR 
in Gubser flow yield similar evolution equation for $A(\omega)$ \cite{Behtash:2017wqg}
\begin{align}
3\omega(\coth^2\rho-1-\mathcal{A}(\omega)) \mathcal{A} ^\prime(\omega)
+\frac{4}{3}(3\mathcal{A}(\omega)+2)^2
+\left(3 \mathcal{A}(\omega)+2\right)\left[\frac{1}{c\omega}-\frac{10}{7}\right]
-\frac{4}{15}=0.
\end{align}


The qualitative differences between the previously discussed Bjorken flow and Gubser flow were established in a number of studies, although it is worth pointing out that the gradient expansion of Gubser flow was found to diverge factorially similarly to the Bjorken case \cite{Denicol:2018pak}. One of the major distinctions between Gubser flow and the previously discussed case of Bjorken flow is the rich range of behaviors Gubser flow experiences. During its evolution, a system with Gubser dynamics undergoes early time free-streaming, progressing to intermediate time hydrodynamization and finally at late times free streaming \cite{Dash:2020zqx}. \com{The final stage is absent from $1+1$d Bjorken flow, as a system with Gubser flow expands more rapidly than in a Bjorken flow and interactions cannot keep up}. In examining the local entropy current and entropy density for the two flows \cite{Chattopadhyay:2018apf,Chattopadhyay:2018fzy}, it was found that Bjorken entropy eventually saturates, while for Gubser flow, the entropy continues to grow for late times, marking the ever-growing departure from thermal equilibrium of the system. 

{Seeing how Gubser flow has more features closer to real heavy ion collisions compared to Bjorken flow, such as transverse expansion, one can speculate that with further relaxed symmetries and the inclusion of transverse dynamics, the attractor story will more resemble Gubser flow.}

\subsection{Phase transitions}

Since hydrodynamic attractors seem to play a role in simpler models of heavy ion collisions, it would be prudent to study the interplay between the attractor and other known features of the QGP, such as spontaneous symmetry breaking. A step in this direction was recently explored in \cite{Mitra:2020hbj} by studying superfluid Bjorken flow, where a conformal fluid in a MIS framework was coupled to a $U(1)$ scalar field undergoing a phase transition via a temperature dependent mass. In this case, the action reads
\begin{align}
S=\int d^4 x \sqrt{-g}\left[p(T)-\frac{1}{2}D_\mu \Sigma D^\mu \Sigma^\star -V(\Sigma, T)\right],\label{u1action}
\end{align}
where $\Sigma=\rho e^{i\psi}$ is the complex scalar field, $\rho$ is the modulus, $\psi$ is the phase, the gauge covariant derivative is $D_\mu \Sigma =(\nabla_\mu+iA_\mu)\Sigma$, the potential is
\begin{align}
V=\frac{1}{2}m^2(T)\rho^2 + \frac{\lambda}{4} \rho^4,
\end{align}
and the metric is the usual Bjorken metric \eqref{bjorken-metric}. The phase transition is characterized by parameterizing the temperature-dependent mass via ${m^2 \sim  (T - T_c)}$. The complete set of equations arise from the energy momentum tensor conservation, the conservation of the $U(1)$ current,
 the scalar equation of motion and the MIS equations. Note that the energy momentum tensor and the current can be decomposed into a sum of normal and superfluid components, e.g.
\begin{align}
j^\mu=\frac{1}{\sqrt{-g}}\frac{\delta S}{\delta A_\mu}=j_n^\mu+j_\psi^\mu,
\end{align}
where $j^\mu_n=nu^\mu$, $n$ is the number density and $j^\mu_\psi= \rho^2 D^\mu\psi.$

As a momentary aside, it is worthwhile to show that the effective fluid action \cite{Brown:1992kc}
\begin{align}\label{fluid-action}
S[g_\mn]=\int d^4 x \sqrt{-g}\; p(T),
\end{align}
indeed generates the hydrodynamic equations via diffeomorphism invariance. 
We define the temperature as
\begin{align}\label{temp}
T\equiv (-\beta^\mu g_\mn \beta^\nu)^{-1/2},
\end{align}
$\beta^\mu=\frac{1}{T}u^\mu$ and the four-velocity, $u^\mu$, is timelike normalized, $u^\mu u_\mu=-1$. 
If we vary \eqref{fluid-action}, using the thermodynamic identity involving the entropy, $s=\frac{\partial p}{\partial T},$ and $\varepsilon+p=sT$, we find 
\begin{align}
\delta S&=\int \ddx \big[\sqrt{-g}\frac{\partial p}{\partial T}\delta T+\delta\sqrt{-g}p(T)\big],\\
&=\int \ddx \sqrt{-g}\big[\frac{1}{2}s T^3\beta^\mu \beta^\nu +\frac{1}{2}p(T)g^\mn\big]\delta g_\mn,\\
&=\frac{1}{2}\int \ddx \sqrt{-g}\big[s T u^\mu u^\nu+p(T)g^\mn \big]\delta g_\mn,
\end{align}
where we recognize the ideal fluid energy momentum tensor in the last line
\begin{align}
T^\mn=(\varepsilon+p) u^\mu u^\nu +p g^\mn.
\end{align}
Under diffeomorphisms $x^\mu\rightarrow x^\mu+\xi^\mu$, we have that 
\begin{align}
\delta g_\mn=\nabla_\mu \xi_\nu+\nabla_\nu\xi_\mu,
\end{align} 
which directly leads to the conservation of the stress tensor after partial integration, $\nabla_\mu T^\mn=0.$ Since we recover the usual ideal hydrodynamic equations from the variation of \eqref{fluid-action}, we will refer to \eqref{fluid-action} as the fluid action.

Dissipation is added to the energy momentum tensor, $T^\mn=T_{\rm ideal}^\mn+\pi^\mn$, and the current $j^\mu=j^\mu_{\rm ideal}+q^\mu,$ with relaxation akin to the MIS formulation,
\begin{align}
\left[\tau_\pi u^\alpha \nabla_\alpha +1\right]\pi^\mn&=-2\eta \sigma^\mn-\zeta \Delta^\mn \nabla_\alpha u^\alpha,\\
\left[\tau_q u^\alpha \nabla_\alpha +1\right](q^\mu+j^\mu_\psi)&=-\kappa\nabla^\mu \frac{\mu}{T},
\end{align}
where $\tau_\pi$ and $\tau_q$ are the relaxation times. Furthermore, the equation for the condensate, $\rho$, and the phase, $\psi$, also have added noise
\begin{align}
-\nabla_\mu \nabla^\mu \rho+\frac{\partial V}{\partial \rho}+\rho D_\mu \psi D^\mu \psi&=\theta_1,\\
\nabla_\mu j^\mu_\psi&=\theta_2.
\end{align}

\begin{figure}\centering
\includegraphics[width=0.75\linewidth]{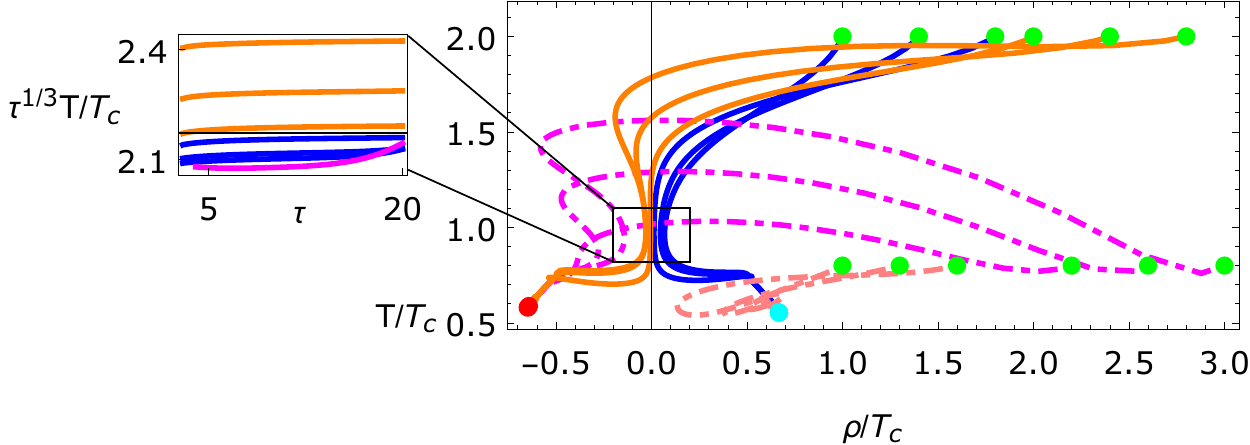}
\caption{Evolution of the system for a variety of initial conditions, which are shown in green. The inset shows the approach of the system to the usual hydrodynamic attractor for certain initial conditions, before veering away and approaching one of two possible fixed points at the end of the system's evolution, shown in red and cyan.  Figure from \cite{Mitra:2020hbj}.}
\label{fig:ssb1}
\end{figure}

We now determine the form of the dissipation by requiring positive entropy production. The entropy is given by $s=(\varepsilon +P-\mu N)/T$, where $\mu$ is the chemical potential. The divergence of the entropy current reads
\begin{align}
\partial_\mu (s u^\mu-\frac{\mu}{T}(q^\mu+j^\mu_\psi))=&-\frac{1}{T}\Pi^\mn\nabla_\mu u_\nu
-((q+j_\psi)\cdot\nabla)\frac{\mu}{T}\nonumber\\
&-\frac{1}{T}\theta_1 (u\cdot \nabla)\rho
-\frac{1}{T}\theta_2(u\cdot D)\psi.
\end{align}
Enforcing positive entropy production leads to the following identification of dissipative structures
\begin{align}
\Pi^\mn=-2\eta\sigma^\mn-\zeta \Delta^\mn \nabla_\alpha u^\alpha,\quad q^\mu+j^\mu_\psi=-\kappa\nabla^\mu \frac{\mu}{T},\\
\theta_1=-\kappa_1 u^\alpha \nabla_\alpha \rho, \quad \theta_2=-\kappa_2 u^\alpha D_\alpha\psi,
\end{align}
where $\eta,$ $\zeta,$ $\kappa,$ $\kappa_1$ and $\kappa_2$ are positive functions of $T$ and $\mu$.

For a conformal system, $p\sim T^4$, the bulk viscosity vanishes, $\zeta=0$, and the transport coefficients can be parameterized via 
\begin{align}
\eta=\frac{4}{3}C_\eta T^3, \quad \tau_\pi = \frac{C_{\tau_\pi}}{T}, \quad \kappa_1= C_{\kappa_1}T, \quad \kappa_2=C_{\kappa_2}T^3.
\end{align}
Defining the pressure anisotropy $\chi:=-3\pi^\eta_\eta/(4T^4)$ and setting $\psi^\prime=0$ as well as $\mu=0,$ for simplicity, the equations of motion are 
\begin{align}
\ddot\rho+\frac{\dot{\rho}}{\tau}+\lambda \rho^3+\sigma(T-T_c)
&=-C_{\kappa_1}T\dot{\rho},\\
\frac{\tau \dot{T}}{T}+\frac{1}{3}(1-\chi)-\sigma \frac{\rho^2+2\tau\rho \dot{\rho}}{8T^3}&=\frac{\tau}{4T^3}C_{\kappa_1}\dot{\rho}^2,\\
\tau \dot{\chi}+\frac{4}{3}\left(\chi-\frac{C_\eta}{C_{\tau_\pi}}\right)+4\chi \tau \frac{\dot{T}}{T}+\frac{\tau T}{C_{\tau_\pi}}\chi&=0.
\end{align}

The relatively simple system described by \eqref{u1action} and the above equations turns out to have a rich set of behaviors. Starting in the unbroken phase with $T>T_c$, the system approaches the usual hydrodynamic attractor ($T\propto \tau^{-1/3}$), as can be seen in Fig.~\ref{fig:ssb1}. During this time, the condensate, $\rho\equiv\vert \Sigma\vert$, value is (close to) zero. Then, some time later, the system reheats and exponentially quickly develops a non-zero value for the condensate. This corresponds to a symmetry-breaking fixed point of the system. The fixed points at $(\pm\rho_\star,t_\star\equiv T/T_c)$ can be computed directly from the equations of motion by recognizing that there should be no dissipation at the fixed point, i.e. $\chi=0$, and $\dot{\rho}=0,$ which means the equation of motion for $\rho$ leads to
\begin{align}
\frac{\partial V}{\partial \rho}=0 \rightarrow \rho_\star=\sqrt{\frac{\sigma}{\lambda}T_c(1-t_\star)}.
\end{align}
Plugging this into the equation for the temperature results in 
\begin{align}
1-t_\star=\frac{8\lambda T_c^2}{3 \sigma^2}t_\star^3,
\end{align}
which in the range $1\geq t_\star \geq 0$ always has a real solution, irrespective of the choice of constants. 
 
Of course, to make contact to phase transitions found in the evolution of the QGP, like the chiral phase transition, one would need to include the full $SU(2)_L\times SU(2)_R\rightarrow SU(2)_V$ symmetry breaking pattern, following the set-up studied in \cite{Grossi:2021gqi}. As such, this study serves as a playground for exploring further interplay between hydrodynamic attractors and other features relevant in the QGP.

\subsection{Attractor of hybrid models}\label{sec:hybrid}

The QGP has a rich structure of strongly and weakly coupled degrees of freedom, interacting in a range of energies. To study the interplay between these two sectors, as a first step it would be worthwhile to study 
 a hybrid fluid model in a Bjorken background \cite{Mitra:2020mei}, where the two fluids represent the hydrodynamic modes of the hard and soft sectors.  The central idea is to see the effect of coupling two viscous fluids which naturally exhibit attractor behavior to one another. A priori, it is not clear how the full system will behave or whether a new attractor will emerge. The discussion here will deviate from \cite{Mitra:2020mei}, but arrive at the same coupling equations studied in the Minkowski background of \cite{Kurkela:2018dku}.

We can study the coupling of two fluids via the fluid action discussed previously.
The action of the first fluid is given by \eqref{fluid-action}, while the second fluid is
\begin{align}\label{fluid-action2}
\tilde{S}[\gt_\mn]=\int \ddx \sqrt{-\gt}\; \pt(T_2).
\end{align}
What is the combined pressure of these two subsystems, \eqref{fluid-action} and \eqref{fluid-action2}, if they are brought in contact? Taking inspiration from Dalton's law \cite{dalton_law}, which states that the total pressure is given by the sum of partial pressures for fixed volume and temperature, i.e. 
\begin{align}\label{dalton}
P=\sum_i P_i,
\end{align}
as well as Agamat's law \cite{bejan_2006}, 
which states that the total volume is given by the sum of partial volumes with fixed pressure and temperature, i.e. 
\begin{align}\label{agamat}
V=\sum_i V_i,
\end{align}
we make the ansatz that the total action is
\begin{align}\label{action-ansatz}
\mathcal{S}[\gb_\mn, T]&=S[g_\mn,T_1]+\tilde{S}[\gt_\mn,T_2],\\
&=\int \ddx \Big{[}\sqrt{-g} P(T_1)+\sqrt{-\gt} \tilde{P}(T_2)\Big{]},
\end{align}
where the combined action is given by the total pressure
\begin{align}
\mathcal{S}=\int \ddx \sqrt{-\gb}\mathcal{P}(T).
\end{align} 
Note that if the effective metrics, $g_\mn$ and $\gt_\mn$, are the same as the physical background metric, $\gb_\mn$, we then have that the total pressure as just the sum of the individual pressures, which is just a statement of Dalton's law \eqref{dalton}. If the pressures are constant and equal, then the integral of the volume elements can be identified as the partial volumes (\ie $\int d^d x\sqrt{-\tilde{g}}=\tilde{V}$), whose sum is the total volume of the system, which is Agamat's law \eqref{agamat}.

As a perhaps useful aside, there seem to be too many metrics. As will be shown later, the effective metrics, $g_\mn$ and $\tilde{g}_\mn$, encode the interaction between the two subsystems. The physical background metric, $\gb_\mn$, is the metric that could be made dynamical \textit{{\`a} la} \cite{PhysRev.94.1468} via
\begin{align}
S=\int d^d x\sqrt{-\gb}(R[\gb_\mn]+\mathcal{P}(T)).
\end{align}
Here, we will consider the background metric $\gb_\mn$ to not have any dynamics, ultimately choosing the Minkowski metric as the background metric, $\gb_\mn=\eta_\mn$.

We will assume that each subsystem is in local thermal equilibrium with a temperature (Killing) vector $\tilde{\beta}^\mu=\frac{1}{T_2}\tilde{u}^\mu$ \cite{Israel:1979wp}, which means the temperature is given by $T_2
=(-\tilde{\beta}^\mu \gt_\mn \tilde{\beta}^\nu)^{-1/2}$. This leads to the following identification when we are in hydrostatic equilibrium \cite{tolman}
\begin{align}\label{equal}
T\sqrt{-\gb_{00}}=T_1 \sqrt{-g_{00}}=T_2 \sqrt{-\gt_{00}}=const.
\end{align}
We will refer to $T$ as the physical temperature, while $T_1$ and $T_2$ will be referred to as the effective temperatures of each subsystem.

We can use the thermodynamic identity $\text{d}\mathcal{P}=\mathcal{S} \text{d}T$ (and also $\text{d}P=s_1 \text{d}T_1$ and $\text{d}\tilde{P}=s_2 \text{d}T_2$) to compute the total entropy density. We thus have
\begin{align}
&\frac{d\mathcal{P}}{dT}=\frac{\sqrt{-g}}{\sqrt{-\gb}}\frac{d P }{d T_1}\frac{dT_1}{dT}+\frac{\sqrt{-\gt}}{\sqrt{-\gb}}\frac{d \tilde{P}}{d T_2}\frac{dT_2}{dT},\\\label{entropy}
&\Rightarrow \mathcal{S}\sqrt{-\gb}=\frac{\sqrt{-g}}{\sqrt{-g_{00}}}s_1+\frac{\sqrt{-\gt}}{\sqrt{-\gt_{00}}}s_2,
\end{align}
which agrees with the result of the entropy in \cite{Kurkela:2018dku}. Thus, the entropy is essentially additive (upto interactions hidden in the effective metric determinants).

Finally, we point out that the two fluids are coupled only through the effective metrics, \ie varying with respect to the effective metrics, $g_\mn$ and $\gt_\mn,$ the microscopic detail is given by
\begin{align}
\nabla_\mu t^{\mu}_{\phantom{\nu}\nu}=0,  \quad \tilde{\nabla}_\mu \tilde{t}^\mu_{\phantom{\mu}\nu}=0,
\end{align}
where the covariant derivative is defined with respect to each subsystem, \ie
\begin{align}\label{emt-cons}
0=\nabla_\mu t^\mn=\partial_\mu t^\mn
+\Gamma^\mu_{\phantom{\mu}\mu\alpha}[g_{\rho\sigma}]t^{\alpha\nu}
+\Gamma^\nu_{\phantom{\mu}\mu\alpha}[g_{\rho\sigma}] t^{\alpha\mu},
\end{align}
and similarly for the tilded equations.
Thus, one fluid only feels the other through the effective metric. In other words, the interactions are encoded by the effective metrics.

We can now compute the full energy-momentum tensor of \eqref{action-ansatz}, where we find
\begin{align}\label{emt-def}
\mathcal{T}^\mn=\frac{\sqrt{-g}}{\sqrt{-\gb}}t^{\alpha\beta}\frac{\partial g_{\alpha\beta}}{\partial \gb_\mn}
+\frac{\sqrt{-\gt}}{\sqrt{-\gb}}\tilde{t}^{\alpha\beta}\frac{\partial \gt_{\alpha\beta}}{\partial \gb_\mn}.
\end{align}
We see that the change in the effective metrics, $g_\mn$ and $\tilde{g}_\mn,$ encompasses the interactions between the two fluids.

We introduce interactions by deforming the effective metrics in such a way that the total energy momentum tensor, $T^\mn$, is conserved in the physical background, $\gb_\mn$, \ie $\nabla_\mu^{(B)} T^\mn=0.$ 
First, we mention that it is straightforward to verify that \eqref{emt-cons} can be rewritten as
\begin{align}\label{ident1}
\partial_\mu (t^\mu_{\phantom{\mu}\nu}\sqrt{-g})&=\frac{1}{2} \sqrt{-g}t^\ab \partial_\nu g_\ab,
\end{align}
and similarly for the other subsystem. Then, using the previous identity, we see that the two subsystems are coupled via the lowest order coupling rules 
\begin{align}\label{couplingeqs}
g_\mn&=\gB_\mn+\frac{\sqrt{-\gt}}{\sqrt{-\gB}}\Bigl{[}\gamma \gB_{\alpha\mu}\tilde{t}^{\alpha\beta}\gB_{\beta\nu}+\gamma^\prime \tilde{t} \gB_\mn \Bigr{]},\nonumber\\
\gt_\mn&=\gB_\mn+\frac{\sqrt{-g}}{\sqrt{-\gB}}\Bigl{[}\gamma\gB_{\alpha\mu} t^{\alpha\beta}\gB_{\beta\nu}+\gamma^\prime t \gB_\mn \Bigr{]},
\end{align}
where we use the shorthand $t=t^\mn \gB_\mn$ and $\tilde{t}=\tilde{t}_\mn \gb^\mn$. Note that the tensor coupling, $\gamma$, and trace coupling, $\gamma^\prime$, have mass dimension $-4$. Higher order couplings, like $\mathcal{O}(\gamma^2,\gamma \gamma^\prime, (\gamma^\prime)^2$), can be included in a straightforward manner, see \cite{Kurkela:2018dku} for a general discussion.
Using \eqref{ident1} and \eqref{couplingeqs}, we can read off the full energy momentum tensor which is conserved in the background metric:
\begin{align}
T^\mu_{\phantom{\mu}\alpha}\sqrt{-\gB}=\frac{1}{2}\left[
(t^\mu_{\phantom{\mu}\alpha}+t_\alpha^{\phantom{\mu}\mu})\sqrt{-g}
+(\tilde{t}^\mu_{\phantom{\mu}\alpha}+\tilde{t}_\alpha^{\phantom{\mu}\mu})\sqrt{-\gt}
- \delta^\mu_\alpha \frac{\sqrt{-g}\sqrt{-\gt}}{\sqrt{-\gB}}\left(\gamma t\cdot \tilde{t}+\gamma^\prime t \tilde{t}\right)
\right],
\end{align}
where $t\cdot \tilde{t}=t^\mn \gb_{\nu\alpha}\tilde{t}^\ab \gb_{\beta\mu}$. Note that stress tensors have their indices raised/lowered by their effective metric, \eg $t^\mu_{\phantom{\mu}\nu}=t^{\mu\alpha}g_{\alpha\nu}.$
The coupling rules \eqref{couplingeqs} are chosen in this way to ensure that the full system has a conserved energy-momentum tensor in the physical background, $\gb_\mn=\eta_\mn$, namely
\begin{eqnarray}
\partial_\mu T^\mu_{\phantom{\mu}\nu} = 0.
\end{eqnarray}

\begin{figure}\centering
\includegraphics[width=0.75\linewidth]{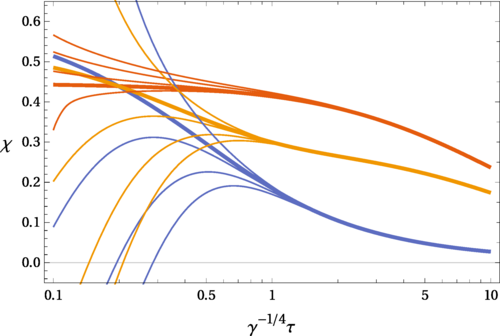}
\caption{Anisotropy as a function of proper time. Note that $\chi=\mathcal{A}/6$, where $\mathcal{A}$ is the anisotropy parameter defined in \eqref{anisotropy-parameter}. The thick lines denote the attractor, while the thin lines denote solutions with different initial conditions. The more viscous system is in red, while the blue lines are the less viscous system. Finally the orange line denotes the emergent system attractor. Figure from \cite{Mitra:2020mei}.}
\label{fig:hybrid}
\end{figure}

To solve the coupling equations \eqref{couplingeqs}, we make the ansatz that the effective metrics will take the diagonal form, motivated by the diagonal form of the background metric, $G_\mn=\eta_\mn$:
\begin{align}\label{metric-ansatz}
g_\mn &=\text{diag}(-a^2,b^2,b^2,c^2),\\
\gt_\mn &=\text{diag}(-\at^2,\bt^2,\bt^2,\ct^2).
\end{align}
Likewise, the two stress tensors are assumed to take the form
\begin{align}
t^\mn&= (\varepsilon+P) u^\mu u^\nu+Pg^\mn+\Pi^\mn
\end{align}
and similarly for the other, tilded subsystem. $\Pi^\mn$ is taken to evolve using the MIS equation \eqref{mis} in each sector with each system having their own shear viscosities, $\eta, \tilde{\eta}$. To distinguish the fluids and in mind to describe the weak/strong coupling sectors found in the QGP, we set the tilded subsystem to be more strongly coupled to have $C_{\tilde{\eta}}=1/4\pi$ to have a smaller shear viscosity than the more weakly coupled untilded sector
\begin{align}
C_\eta&=10 C_{\tilde{\eta}}, \quad C_{\tau}=5 C_\eta,\\
C_{\tilde{\tau}}&=\frac{2-\ln 2}{2\pi}.
\end{align}
Thus the untilded weakly coupled sector can be thought of as governed by kinetic theory, while the tilded strongly coupled sector is governed by a holography.

It is worth noting that in the decoupling limit, $\gamma,\gamma^\prime\rightarrow 0$, each subsystem contains an attractor. As can be seen in Fig.~\ref{fig:hybrid}, with interactions turned on, the combined system surprisingly has an emergent attractor as well. This emergent attractor is parameterized by the energy densities of each subsystem, $\gamma \varepsilon$ and $\gamma \tilde{\varepsilon}$. Interestingly, the scenario modelled here is reminiscent of bottom-up thermalization \cite{Baier:2000sb}, in that at early times the strongly coupled subsystem is dominant, while at later times it is the weakly coupled subsystem which dominates. Intriguingly, though the individual subsystems were taken to be conformal ($t^\mn g_\mn=0=\tilde{t}^\mn \tilde{g}_\mn$), the full system is in general not, $T^\mn G_\mn\neq0,$ becoming conformal at late times.

The hydrodynamization time for each subsystem can be defined (see \cite{Heller:2016rtz,Attems:2017zam} for similar definitions) when the pressure is sufficiently close to first order hydrodynamics, namely
\begin{align}
\frac{\vert\phi-\phi_{\rm 1st}\vert}{P}<0.1, \quad \text{for } \tau>\tau_{\rm hyd},
\end{align}
and similarly for the other sector. It was found in a variety of cases that the hydrodynamization time for the harder sector is longer than for the strongly interacting sector. One can compare the ratio of the hard to soft hydrodynamization times via the ratio $R_{\rm hd}=\tau_{\rm hd}/\tilde{\tau}_{\rm hd}$, where it is clear that depending on the initial conditions, $R_{\rm hd}$ can vary by an order of magnitude, perhaps indicating that the hydrodynamization scenario in small-system collisions \cite{Chesler:2015bba,Loizides:2016tew,Schlichting:2016sqo} may be quite different to their heavier ion counterpart.

The hybrid coupling described in this section lends itself naturally as a framework to study more nuanced examples, closer to the physics of the QGP. For instance, instead of fluids, one can consider the coupling between an ultraviolet sector described by kinetic theory, while the low energy degrees of freedom are described holographically. Such a framework is known as semiholography, a term coined in \cite{Faulkner:2010tq} and introduced to heavy ion collisions in \cite{Iancu:2014ava}. It would be interesting to see the behavior of the attractor with the democratic metric coupling in addition to the scalar coupling between sectors \cite{Banerjee:2017ozx,Ecker:2018ucc,Mondkar:2021qsf}, which drives the energy from the hard to the soft sector.

\section{Holographic attractor}\label{sec:holo}

The connection between hydrodynamic and holographic theories is firmly established, e.g. see \cite{Bhattacharyya:2008jc} for a discussion bridging BRSSS hydrodynamics to a holographic theory. Moreover, boost-invariant holography had been studied by numerous groups over the years \cite{Heller:2011ju,Heller:2012je,Jankowski:2014lna}, going back to the linearized analysis of \cite{Janik:2006gp}. The holographic setting was also where the Borel resummed hydrodynamic gradient expansion was first considered \cite{Heller:2011ju,Heller:2013fn}. As such, after finding the hydrodynamic attractor \cite{Heller:2015dha}, the stage was set to explore the attractor story in the strong coupling context. An important early work in this direction was due to 
Romatschke \cite{Romatschke:2017vte}, whose example we will reproduce and expand on in Sec.~\ref{sec:holoex}.
We will then turn our attention to some of the novel directions explored in holographic attactors.

\subsection{An illustrative example}\label{sec:holoex}
Here we will follow \cite{Romatschke:2017vte}. In the holographic context, one aims to solve a classical gravity problem which is dual to a strongly coupled quantum field theory. We will want this dual theory to be comparable to the hydrodynamic theory. The dual theory is encoded in the boundary conditions of the gravity equations, which we will outline shortly. We begin with the vacuum Einstein equations 
\begin{align}\label{EE}
R_\mn -\frac{1}{2}g_\mn R + \Lambda g_\mn=0,
\end{align} 
where $R_\mn$ is the Ricci tensor and $R\equiv R_\mn g^\mn$ is the Ricci scalar. The cosmological constant in AdS is given by $\Lambda=-{(d-1)(d-2)}/{2L^2}=-6$, since we work in $d=5$ dimensions and with the AdS radius, $L$, set to unity.

The metric in Eddington-Finkelstein coordinates is given by
\begin{align}\label{ef-metric}
ds^2=2 dr d\tau-A d\tau^2+\sigs^2 \left(e^B dx_\perp^2+ e^{-2B}d\xi^2\right),
\end{align}
where $\xi$ is the spacetime rapidity.
We take the functions $A$, $\sigs$ and $B$ to be functions of the proper time, $\tau$, and the holographic radius, $r$. 
With this metric ansatz, the Einstein equations \eqref{EE} can be helpfully written as a set of nested ordinary differential equations \cite{Chesler:2008hg}
\begin{align}\label{ee1}
0&=\sigs\dot{\sigs}^\prime+2\sigs^\prime \dot{\sigs}-2\sigs^2,\\
0&=2\sigs \dot{B}^\prime+3(\sigs^\prime \dot{B}+B^\prime\dot{\sigs}),\label{ee2}\\
0&=A^{\prime\prime}+3B^\prime\dot{B}-12\sigs^\prime\dot{\sigs}\sigs^{-2}+4,\label{ee3}\\
0&=2\ddot{\sigs}+\dot{B}^2\sigs-A^\prime\dot{\sigs}^\prime,\label{const1}\\
0&=2\sigs^{\prime\prime}+B^{\prime2}\sigs,\label{const2}
\end{align}
where the prime denotes the radial derivative, $h^\prime \equiv \partial_r h$, and the dot denotes the radial derivative along the null geodesic
\begin{align}
\dot{h}\equiv \partial_\tau h+\frac{1}{2}A\partial_r h.
\end{align}
Note that \eqref{const1} and \eqref{const2} are constraint equations, while the other equations are to be solved numerically. 

Near the boundary at $r=\infty,$ the asymptotic expansion of the metric functions is given by
\begin{align}
A(\tau,r)&=r^2 \sum_{n=0}^\infty a_n(\tau) r^{-n} ,\\
B(\tau,r)&=\sum_{n=0}^\infty a_n(\tau) r^{-n} ,\\
\sigs(\tau,r)&=r \sum_{n=0}^\infty a_n(\tau) r^{-n}.
\end{align}
We want the metric at the boundary to take the form of the Bjorken metric, \eqref{bjorken-metric}, in order for the dual theory to exhibit boost invariance. 
This implies that the boundary conditions at $r\rightarrow \infty$ are
\begin{align}
A=r^2, \quad B=-\frac{2}{3}\ln {\frac{1+r \tau}{r}} \quad \text{and}\quad \sigs^3 =r^2(1+r\tau).
\end{align}
This determines the leading coefficients in the expansion of the metric functions, $a_0 =1$ and $s_0=1.$ Furthermore, it is easy to see that the metric \eqref{ef-metric} remains invariant under the diffeomorphism, $r\rightarrow r+f(\tau).$ This residual gauge freedom can be fixed by setting $a_1=0.$

The boundary stress tensor 
\begin{align}
T_\mn=\text{diag}(\varepsilon, p_T,p_T, p_L)
\end{align}
is related to the metric coefficient $a_4$ via holographic renormalization \cite{Balasubramanian:1999re,deHaro:2000vlm} by
\begin{align}
\varepsilon&=-\frac{3\kappa}{4}a_4,\\ \label{long-p}
p_L&=-\varepsilon-\tau \partial_\tau \varepsilon,\\
p_T&=\varepsilon+\frac{1}{2}\tau \partial_\tau \varepsilon.\label{tran-p}
\end{align}

The Einstein equations are then solved using numerical techniques 
due to Chesler and Yaffe (namely, the spectral method) \cite{Chesler:2008hg,Chesler:2009cy,Chesler:2013lia}. 
There is also publicly available code in \cite{Wu:2011yd,paul-attractor}.

Various initial conditions at $\tau=\tau_0$ were generated in \cite{Romatschke:2017vte} by taking $\Sigma$ to be
\begin{align}
\Sigma^3=r^2(1+r\tau)+\frac{r^2 s_2+r s_3+s_4}{r^4+\bar{c}},
\end{align}
where 
\begin{align}
s_2&=\frac{1}{20}\left(4\varepsilon +3 \tau \partial_\tau \varepsilon\right),\\
s_3&=\frac{1}{40}\left(13 \partial_\tau \varepsilon+5\tau \partial^2_\tau \varepsilon \right),\\
s_4&=\frac{1}{560\tau}\left(9\partial_\tau \varepsilon+135 \tau \partial_\tau^2\varepsilon+35 \tau^2 \partial_\tau^3\varepsilon \right).
\end{align}
Then choosing the profile $\varepsilon\sim \tau^\alpha$, one can generate initial data by changing $\alpha$ and $\bar{c}.$  

\begin{figure}
\includegraphics[width=\linewidth]{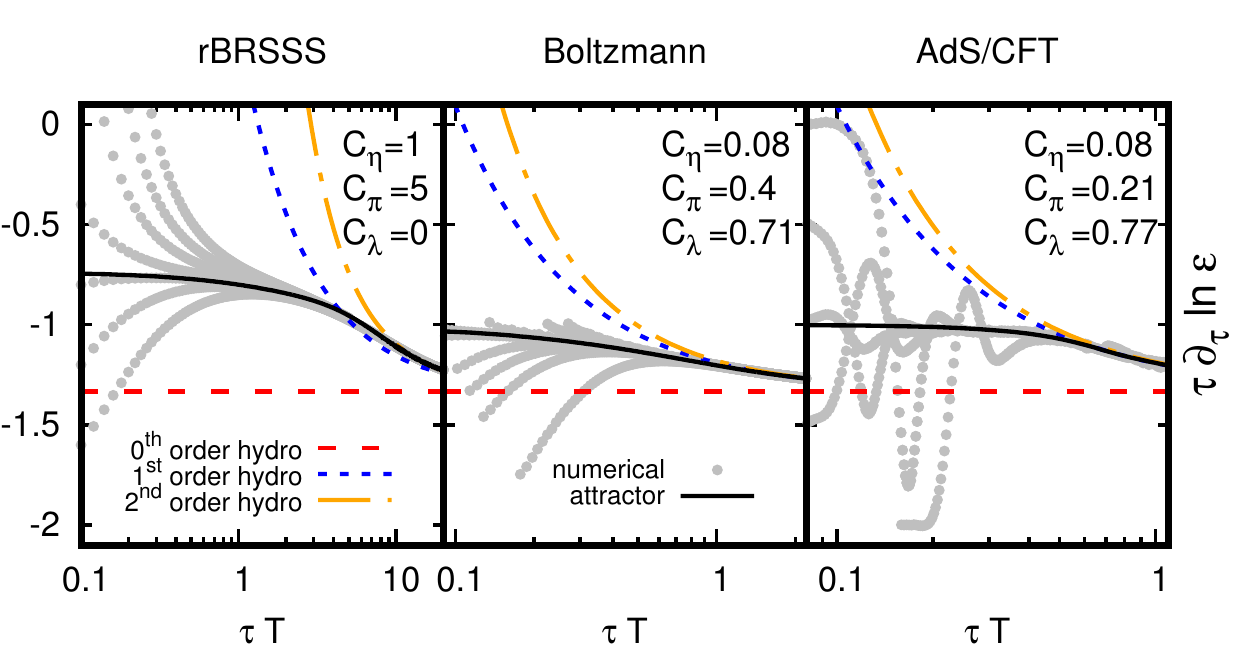}
\caption{The attractor for rBRSSS hydrodynamics (left panel), kinetic theory in the relaxed time approximation (middle panel) and the holographic theory from Sec.~\ref{sec:holoex} (right panel). The gray lines denote solutions with different initial conditions, while the solid lines are the numerical attractor. The late time behavior should approach the hydrodynamics, which are denoted by the dashed lines. Note that the agreement between the attractor and hydrodynamics at late times occur at different timescales, $\omega=\tau T$, for the different theories. Figure from \cite{Romatschke:2017vte}.}
\label{fig:holography}
\end{figure}

Sorting through the initial data, there is a clear late-time attractor as can be seen in Fig.~\ref{fig:holography}. However, at earlier times the story is not so clear 
due to the well-known oscillatory nature of $N=4$ SYM, which arises primarily from the existence of non-hydrodynamic modes at early times \cite{Romatschke:2017vte,Spalinski:2017mel}. 

To identify the holographic attractor, it was posited in \cite{Romatschke:2017vte} that the relevant condition should be that the attractor experiences little oscillatory behavior and whose initial value tends to
\begin{align}\label{romat-attractor}
\lim_{\tau\rightarrow0}\frac{\partial \ln \varepsilon}{\partial \ln \tau}\rightarrow -1.
\end{align}
The attractor is then visible in the rightmost panel of Fig.~\ref{fig:holography} as a solid black line, while the various initial conditions produce a variety of oscillatory behavior. Such an ambiguity is not found in relativistic hydrodynamics and kinetic theory, where the approach to the attractor tends to be monotonic from a given set of initial conditions.

In contrast to the Romatschke approach, another study was undertaken in \cite{Kurkela:2019set}, where the initial conditions of the anisotropic part of the metric was modelled instead in two ways (note: $\rho=1/r$)
\begin{align}
B(\tau_0, \rho)+\frac{2}{3}\log{(\tau_0+\rho)}=
\begin{cases}
 e^{-40\rho T}32 \rho^5 T^5& \text{(UV)},\\
32 \rho^5 T^5& \text{(IR)}.
\end{cases}
\end{align}
This particular choice of initial conditions was chosen such that the initial ratio of the longitudinal pressure to the energy density $p_L(\tau_0)/\varepsilon(\tau_0)$ and its first derivative are equal. The meaningful difference between the two families of initial conditions is in their extent in the holographic direction, with the UV solution localized almost entirely on the boundary. Both sets of initial conditions display the expected oscillatory behavior due to the expected decay of quasinormal modes. Intriguingly, the more localized UV initial condition seems to approach the hydrodynamic behavior sooner than the corresponding IR family of initial conditions. For either family of initial conditions, the approach to the attractor takes longer than the corresponding set-up in kinetic theory or IS hydrodynamics, irrespective of initialization time. Thus, results of \cite{Kurkela:2019set} seem to indicate that the attractor only exists in a meaningful way at late times for holographic set-ups with the early time dynamics dominated by transients, \com{which is further supported by related work of higher-order fluid dynamics \cite{Chattopadhyay:2018pwe}.}

\subsubsection{Borel summation}\label{sec:borel}

Whether the condition \eqref{romat-attractor} is the ``best'' or unambigous identifier for a holographic attractor is up for debate. A perhaps more rigorous approach is due to \cite{Spalinski:2017mel,Spalinski:2018mqg}. Using the same metric as in \eqref{ef-metric}, the hydrodynamic gradient expansion of the energy density in terms of the proper  time
\begin{align}\label{expansion}
\varepsilon=\frac{3N_c^2 \pi^2 }{8}\tau^{-4/3}\sum_{i=2}^{\infty} \varepsilon_i \tau^{-2(i-2)/3}
\end{align} 
was computed up to $240^{th}$ order \cite{Heller:2013fn}. The longitudinal and transverse pressure in the Bjorken case is given by \eqref{long-p} and \eqref{tran-p},
so that one can easily compute the anisotropy parameter \eqref{anisotropy-parameter}, which has an expansion in terms of $\omega=T\tau$:
\begin{align}\label{grad-exp}
A(\omega)=\sum_{k=1}^\infty a_k \omega^{-k},
\end{align}
Thus, the expansion of the energy density in terms of proper time \eqref{expansion} can be re-expressed as an expansion in the anisotropy parameter in terms of the dimensionless time, $\varepsilon(\tau)\leftrightarrow A(\omega)$.  
One then computes the Borel transform of the original power series, which is given by
\begin{align}
\mathcal{B}A(x)=\sum_{k=1}^\infty \frac{a_k }{k!} x^k
\end{align}
The intent of the Borel transform is to provide a method of handling divergent series. The gradient expansion \eqref{grad-exp} is a divergent expansion with larger order terms diverging factorially \cite{Heller:2015dha}. One can then perform the Borel resummation via the inverse Borel transform
\begin{align}\label{resummed}
A(\omega)=\omega \int_C d x e^{-\omega x} \mathcal{B}A(x),
\end{align}
where the contour, $C$, connects $x=0$ to $x=\infty$ in the complex plane. Practically, since only 240 terms are known due to the high technical cost of computing the coefficients, the analytic continuation is done using Pad{\'e} approximants. The Pad{\'e} approximants are, in some respects, a generalization of the Taylor series, with the improvement that singularities are explicitly included in the approximant. This is done by considering a ratio of polynomials, i.e. we can approximate a function via
\begin{align}
f(x)\approx \frac{\sum_{i}^N a_i x^i}{1+\sum_j^M b_j x^j},
\end{align}
where the denominator can better capture the singularity structure of the function that is being approximated, compared to Taylor series, which does not contain such terms. For more on Pad{\'e} approximants, see \cite{numrecipe} and \cite{Ellis:1995jv} for application to QCD sum rules.

Does this method provide a good approximation to the attractor? In the case the attractor is known, such as in the previously mentioned HSJW model \eqref{hsjw}, the known attractor can be compared to the result of the integrating the series truncated at $240^{th}$ order via \eqref{resummed}, where there was quantitative agreement down to about $\omega\approx 0.3,$ indicating the method provides meaningful results. In the case of holography, where the exact attractor is not known, it was found in \cite{Spalinski:2017mel} that there is good agreement with \cite{Romatschke:2017vte} up to $\omega\sim 0.4,$ whereas the qualitative behavior is different for smaller $\omega$.

\subsection{Violation of energy conditions}

As the main workhorse for the study of holographic attractors and the go-to holographic toy model for heavy ion collisions, it is important to understand the holographic Bjorken model in Sec.~\ref{sec:holoex}.
For this reason, it is interesting to note the holographic Bjorken flow can evolve to violate the dominant and weak energy conditions \cite{Rougemont:2021qyk}, even with initial conditions that satisfy these bounds. For conformal Bjorken flow, the weak energy condition was given in the first holographic study of Bjorken flow \cite{Janik:2005zt}
\begin{align}
T^\mn t^\mu t^\nu\geq 0,
\end{align}
where $t^\mu$ is any time-like vector, which implies for Bjorken flow
\begin{align}
\varepsilon(\tau)\geq 0, \quad \varepsilon^\prime(\tau)\leq 0, \quad \text{and} \quad \tau \varepsilon^\prime(\tau)\geq -4 \varepsilon(\tau).
\end{align}
Note that since we are considering a conformal system, the strong energy condition 
\begin{align}
T_\mn t^\mu t^\nu \geq \frac{T^\mu_\mu}{2}
=0,\end{align}
is equivalent to the weak energy condition.
The dominant energy condition implies that 
\begin{align}
\varepsilon(\tau)>0, \quad \text{and} \quad -1<\frac{p_T-p_L}{\varepsilon}< 2.
\end{align}
It was found that the violations occur at times around $\tau\approx 0.5 \text{ fm/c}$, which is within the prehydrodynamic regime in the QGP \cite{Kurkela:2018wud}. This can have implications for the validity of kinetic theory, the typical effective description used in this regime.
Kinetic theory satisfies the energy conditions: the weak energy condition implies that the energy density is non-negative and the dominant energy condition is the requirement that matter does not exceed the speed of light \cite{wald_2009}. \com{If these violations were shown to exist in the energy mometum tensor during nuclear collisions, this could indicate that this transient regime would lie outside of the domain of validity of kinetic theory descriptions \cite{Rougemont:2021qyk}, indicating that strongly-coupled hydrodynamization as predicted by holography would be the preferred description.}

\subsection{Gauss-Bonnet attractor}\label{sec:GB}

Attractors have been studied in Gauss-Bonnet (GB) holography \cite{Casalderrey-Solana:2017zyh,Gushterov:2018erj,Casalderrey-Solana:2019npu}. The key advantage of studying GB holographic systems is that one has access to intermediate coupling (in addition to infinitely strong and weak coupling) regimes. The only parameter is $\lambda_{GB}$, which controls the size of the higher derivative corrections.  The dual field theory is unknown, but finite and negative values of $\lambda_{GB}$ are at least qualitatively similar to what one would expect from finite coupling corrections to $N=4$ SYM. The negative values of $\lambda_{GB}$ do lead to violations of causality in the ultraviolet \cite{Camanho:2014apa}, but the interest of these works is in the direction of the infrared dynamics and the system's approach to hydrodynamics.

Here, we will briefly outline some details of the construction and the lessons learnt.
The GB action is 
\begin{align}
S=\frac{1}{2\kappa_5^2}\int \text{d}^5 x \sqrt{-g}\left[R+\frac{12}{L^2}+\frac{\lambda_{GB}L^2}{2}(R_{\mu\nu\rho\sigma}R^{\mu\nu\rho\sigma}-4R_\mn R^\mn+R^2)\right],
\end{align}
where $\kappa_5$ is related to Newton's constant in 5 dimensions and $L$ is the AdS radius for the $\lambda_{GB}=0$ theory. Note that the $\lambda_{GB}=0$ case corresponds to the example in Sec.~\ref{sec:holoex} with the terms proportional to $\lambda_{GB}$ representing higher derivatives of the metric. The shear viscosity of this theory is given by
\begin{align}
\frac{\eta}{s}=\frac{1-4\lambda_{GB}}{4\pi}.
\end{align}
Using a similar metric ansatz of the Eddington-Finkelstein form \eqref{ef-metric}, 
one then solves the Einstein equations as outlined in Sec.~\ref{sec:holoex}.

\begin{figure}\centering
\includegraphics[width=0.75\linewidth]{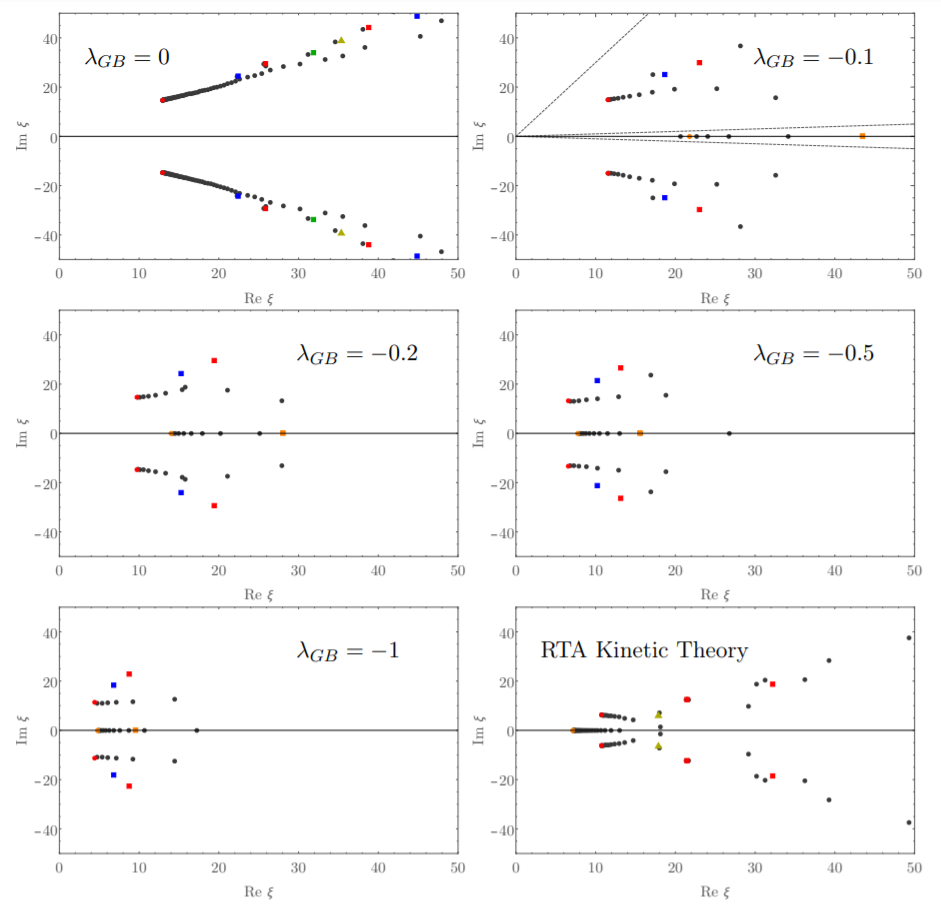}
\caption{Positions of poles of the Pad{\'e} approximant in the complex $\chi$ plane (which we call $x$ in Sec.~\ref{sec:borel}) for different values of the coupling $\lambda_{GB}$, compared to the RTA kinetic theory result. Note that $\lambda_{GB}=0$ represents the typical pole structure seen in holography. Figure from \cite{Casalderrey-Solana:2017zyh}.}
\label{fig:gb}
\end{figure}

For the values of $\lambda_{GB}$ studied in \cite{Casalderrey-Solana:2017zyh}, the attractor is estimated using Borel summation as outlined in Sec.~\ref{sec:borel}. It was found that the relaxation time of the non-hydrodynamic modes occurs when the gradients of the system are still large, i.e. for anisotropy parameter $>1$, with first order viscous hydrodynamics comparable to the estimated attractor.
Interestingly, it is found that by varying $\lambda_{GB}$ there is a qualitative interpolation in the singularities found in the complex plane between the holographic infinite coupling limit and the weakly coupled limit in kinetic theory, which can be inferred from Fig.~\ref{fig:gb}. 

\subsection{$D$-dimensional Bjorken flow}

The attractor in arbitrary $D=n+1$ dimensions was studied in \cite{Casalderrey-Solana:2018uag}. For a review of large $D$ gravity, see \cite{Emparan:2020inr}. The essential technical difference is that the metric \eqref{ef-metric} has the same form and dependence, but there are more transverse directions, i.e.
\begin{align}
ds^2=2 dr d\tau-A d\tau^2+\sigs^2 \left(e^B dx_\perp^2+ e^{-(n-2)B}d\xi^2\right),
\end{align}
which is chosen such that the determinant of the term in brackets is equal to one. Operationally, the $D-$dimensional Einstein equations are rewritten in the same way as in \eqref{ee1}-\eqref{const2}. The equations are then expanded in a large $n=D-1$ limit up to $n^{-3}$ order.
The key ingredient to compare holographic theories in different dimensions is the requirement that the late time behavior is the same, which is enforced by introducing the late time scale, $\Lambda,$ which arises from considering the late time behavior of the ideal Bjorken expansion
\begin{align} 
\varepsilon_{\rm ideal}=\frac{\Lambda^n}{(\Lambda \tau)^{n/(n-1)}}.
\end{align}

Interestingly in the case studied in \cite{Casalderrey-Solana:2018uag}, at the time scale $\tau\sim 1/\Lambda$  
the decoupling of non-hydrodynamic modes from the hydrodynamic modes occurs in an average sense: these modes are rapidly oscillating around the hydrodynamized evolution, averaging to zero over the typical expansion time scale. Coupled with the fact that the non-hydrodynamic excitations have amplitudes which are not parametrically suppressed, it seems like the typical approach to the attractor would be different to the usual $D=5$ case.

\section{Attractor in kinetic theory}\label{sec:kt}

 In heavy ion collisions, the conceptual picture is that kinetic theory tracks the evolution of (quasi-)particles. The applicability of the theory is early times post collision up to the time when the build-up of the soft gluon bath is significant \cite{Busza:2018rrf}. This corresponds to roughly the hydrodynamization time, $\tau\sim 1 \text{ fm/c}$. The bridging between regimes where the dominant effective description is kinetic theory to hydrodynamics has been developed in K{\o}MP{\o}ST \cite{Kurkela:2018wud,Kurkela:2018vqr}.
 
As a workable theory when the heavy ion system is quite far-from-equilibrium, it is interesting to study which role attractors play in a kinetic description. To that end, we present the reader with an example kinetic computation in Sec.~\ref{sec:kinex}, before describing further progress in the following subsections. 

\subsection{Kinetic theory primer}\label{sec:kinex}

Here we provide a short recap of key concepts regarding kinetic theory, following the discussion in \cite{Gorbunov:2011zzc}. Kinetic theory concerns itself with the microscopic detail of particles directly. The fundamental object of interest in kinetic theory, keeping track of the particles' positions and momenta, is the one
particle distribution function, $f=f(x^\mu,p_\mu)$, where $x^\mu$ are the spacetime coordinates and $p_\mu$ is the four-momentum. Note that the distribution function should not be confused with the dimensionless quantity \eqref{fdef}, used to characterize the hydrodynamic attractor.
The evolution of the distribution function in phase space follows Liouville's theorem \cite{landau}, namely 
\begin{align}
0=\frac{df}{d\lambda }=\frac{d x^\mu}{d \lambda}\partial_\mu f+\frac{dp_\mu}{d\lambda} \frac{\partial f}{\partial p_\mu}
\end{align}
where $\lambda$ is an affine parameter. This equation is commonly known as the collisionless Boltzmann equation. It is common to consider the Boltzmann equation on-shell, e.g. $p_\mu p^\mu =-m^2$ in our sign convention. This means that the one of the components of the four-momentum is determined in terms of the other momenta with a common choice being $p_0=p_0(p_i).$ For a massive gas in a Minkowski background, this is simply $p_0=\sqrt{p_i \delta^{ij}p_j+m^2}.$ 

Macroscopic quantities can be computed straightforwardly by considering the moments of the distrbution functions, i.e. integrating the distribution function with increasing powers of the four-momentum over momentum. For instance, the particle
 number current and energy momentum tensor are given by \cite{Andreasson:2011ng,Gorbunov:2011zzc}
\begin{align}
n^\mu(x^\mu)&=\int \frac{d^3 p}{(2\pi)^3\sqrt{-g}}\frac{p^\mu }{p^0} f(x^\mu,p_\mu),\\
\label{kt-emt}
T^{\mu\nu}(x^\mu)&=\int \frac{d^3 p}{(2\pi)^3\sqrt{-g}}\frac{p^\mu p^\nu}{p^0} f(x^\mu,p_\mu),
\end{align}
respectively.

Noting that the four-momentum is $p^\mu=\frac{dx^\mu}{d\lambda}$ and the four-force is $F_\mu=\frac{dp_\mu}{d\lambda}$, we find that the distribution function evolves according to the Boltzmann equation in a curved spacetime in the absence of external forces
\begin{align}\label{boltzmann}
p^\mu \partial_\mu f+p_\mu p^\beta\Gamma^\mu_{\phantom{\mu}\alpha\beta}\frac{\partial f}{\partial p_\alpha}=C[f],
\end{align}
where $C[f]$ is the collision kernel, which encompasses all order scattering processes. In practice, it is more expedient to work with a simpler, truncated collision kernel. 
 For analytic study, a helpful choice of the collision kernel is the relaxation time approximation (RTA) (also known as the Bhatnagar-Gross-Krook (BGK) approximation \cite{Bhatnagar:1954zz}), where the collision term is given by
\begin{align}\label{rta}
C[f]=\frac{p^\mu u_\mu}{\tau_R}(f-f_{eq}),
\end{align}
where $\tau_R$ is the relaxation time, which in principle can depend on the proper time, $\tau_R=\tau_R (\tau)$. Here, we will take the equilibrium distribution function to be given by
\begin{align}
f_{eq}=e^{p^\mu u_\mu /T}
\end{align}
where $T$ is the temperature and will be a function of proper time, $\tau$.
The interpretation of the RTA collision term is that the system relaxes back to its equilibrium value, which can be easily seen in the following example. Assume that we are in flat Minkowski spacetime with only time dependence in the distribution function. Taking $f_{\rm eq}$ to be a constant and shifting $f-f_{\rm eq}\rightarrow f$ to write the Boltzmann equation \eqref{boltzmann} in the RTA \eqref{rta} as
\begin{align}
\partial_0 f=-\frac{1}{\tau_R}f,
\end{align}
which has solution after shifting back as
\begin{align}
f=f_{\rm eq}+e^{-t/\tau_R}.
\end{align}
Thus, we see that the distribution function exponentially rapidly approaches its equilibrium value in a timescale, $\tau_R.$

An important ingredient in the RTA is what is known as the matching condition. Since the energy momentum tensor \eqref{kt-emt} is conserved, $\nabla_\mu T^\mn=0,$  the first moment of the Boltzmann equation must vanish. This sets a condition on the collision kernel, namely that the moment of the collision kernel should vanish, leading to the dynamical matching conditions. In other words, the energy density, $\varepsilon$, should match the equilibrium energy density, $\varepsilon_{\rm eq}$, i.e.
\begin{align}
T^{00}&=T^{00}_{\rm eq},\\
&\Rightarrow \int \frac{d^3 p}{(2\pi)^3} \frac{(p^0)^2}{p^0}(f-f_{\rm eq})=0.
\end{align}

\subsubsection{Attractor via the method of moments}

A straightforward way of seeing the attractor in kinetic theory is using the method of moments. In the context of attractors, the tower of moments was studied in the RTA \cite{Kurkela:2019set,Strickland:2019hff} as well as with more realistic collision kernels \cite{Almaalol:2020rnu}, following prior work 
in studying the divergence of the Chapman-Enskog expansion \cite{Denicol:2016bjh}. Moreover, for an application of studying the fixed points, e.g. free streaming and hydrodynamic, found in kinetic theory via the method of moments, see \cite{Blaizot:2017ucy,Blaizot:2019scw,Blaizot:2021cdv}. Here we outline the basics of the construction.

Instead of working with the distribution function directly, it is helpful to consider the generalized moments, $\rho$, of the distribution function 
\cite{Denicol:2016bjh}
\begin{align}\label{moments}
\rho_{n,l}=\int \frac{d^3 k}{(2\pi)^3\tau}(k^0)^n\left(\frac{k_\eta}{k^0\tau}\right)^{2l} f(\tau,k_0,k_\eta).
\end{align} 
The moments contain all of the microscopic information of the distribution function. For instance, 
it is clear by inspection that the lower moments have a straightforward physical interpretation, namely the energy density is given by $\varepsilon=\rho_{1,0}$ and the longitudinal momentum is given by $p_L =\rho_{1,1}$. 

The evolution equation for the moments is found by integrating the Boltzmann equation.
One sees that 
\begin{align}
\int \frac{d^3 k}{(2\pi)^3\tau}(k^0)^n\left(\frac{k_\eta}{k^0\tau}\right)^{2l}\partial_\tau f&=\partial_\tau \rho_{n,l}-\int \frac{d^3 k}{(2\pi)^3}k_\eta^{2l}\partial_\tau\left[ \tau^{-1-2l}(k^0)^{n-2l}\right]f,\\
&=\partial_\tau \rho_{n,l}+\frac{1+2l}{\tau}\rho_{n,l}+\frac{n-2l}{\tau}\rho_{n,l+1},
\end{align}
where we integrated by parts and used that in the Bjorken massless case $k_0=\sqrt{k_x^2+k_y^2+k_\eta^2/\tau^2}$. Integrating over the equilibrium distribution function leads to
\begin{align}
\rho^{\rm eq}_{n,l}=\int \frac{d^3 k}{(2\pi)^3}(k^0)^n \left(\frac{k_\eta}{\tau k^0}\right)^{2m} f_{\rm eq}.
\end{align}
Putting together the pieces, we find that the general equation for moments is given by
\begin{align}
\partial_\tau \rho_{n,l}+\frac{1+2l}{\tau} \rho_{n,l}+\frac{n-2l}{\tau}\rho_{n,l+1}=-\frac{\rho_{n,l}-\rho^{\rm eq}_{nm}}{\tau_R}.
\end{align} 

In the case considered by \cite{Kurkela:2019set}, namely the set of lowest lying equations $(n,l)=(1,0)$ and $(n,l)=(1,1)$, one finds the following set of equations
\begin{align}\label{moment1}
0&=\partial_\tau \varepsilon+\frac{1}{\tau}(\varepsilon+p_1),\\
0&=\partial_\tau p_1+\frac{1}{\tau}(3p_1-p_2)+\frac{p_1-\varepsilon/3}{\tau_R}.\label{moment2}
\end{align}
As a reminder, the relaxation time is in principle a function of proper time. As above, we will consider the conformal case, where $\tau_R^{-1}\propto \varepsilon^{1/4}.$
We will use the rescaled time $t\equiv \tau/\tau_R$ (not to be confused with the Minkowski time, $t$). A helpful identity to be used below is that
\begin{align}
\frac{1}{\tau_R}\frac{d t}{d \tau}=\frac{1}{\varepsilon^{1/4}}\frac{d(\tau \varepsilon^{1/4})}{d\tau}=1+\frac{\tau}{4}\partial_\tau \varepsilon=\frac{1}{4}(3-x), 
\end{align} where we introduced $x\equiv p_1/\varepsilon$. Finally, calling $y(t)\equiv p_2/\varepsilon$, we see that \eqref{moment1} and \eqref{moment2} can be combined in one equation as \cite{Kurkela:2019set}
\begin{align}\label{moment-att}
\frac{3}{4}\left(3-x\right)t x^\prime (t)=3x^2-6x+3y+t(1-3x).
\end{align}
In analyzing \eqref{moment-att}, a useful starting point is to consider is at $t=0.$ Requiring the derivative of $x$ is finite at $t=0,$ we see that we have two cases
\begin{align}
x(0)=1\pm \sqrt{1-y(0)}.
\end{align}
At late times, it is clear that \eqref{moment-att} becomes
\begin{align}
\frac{3}{4}(3-x)x^\prime (t)=1-3x,
\end{align}
from which it follows that the generic late-time behavior is $ \lim_{t\rightarrow \infty}x(t)\rightarrow \frac{1}{3}.$ 

\begin{figure}\centering
\includegraphics[width=0.7\linewidth]{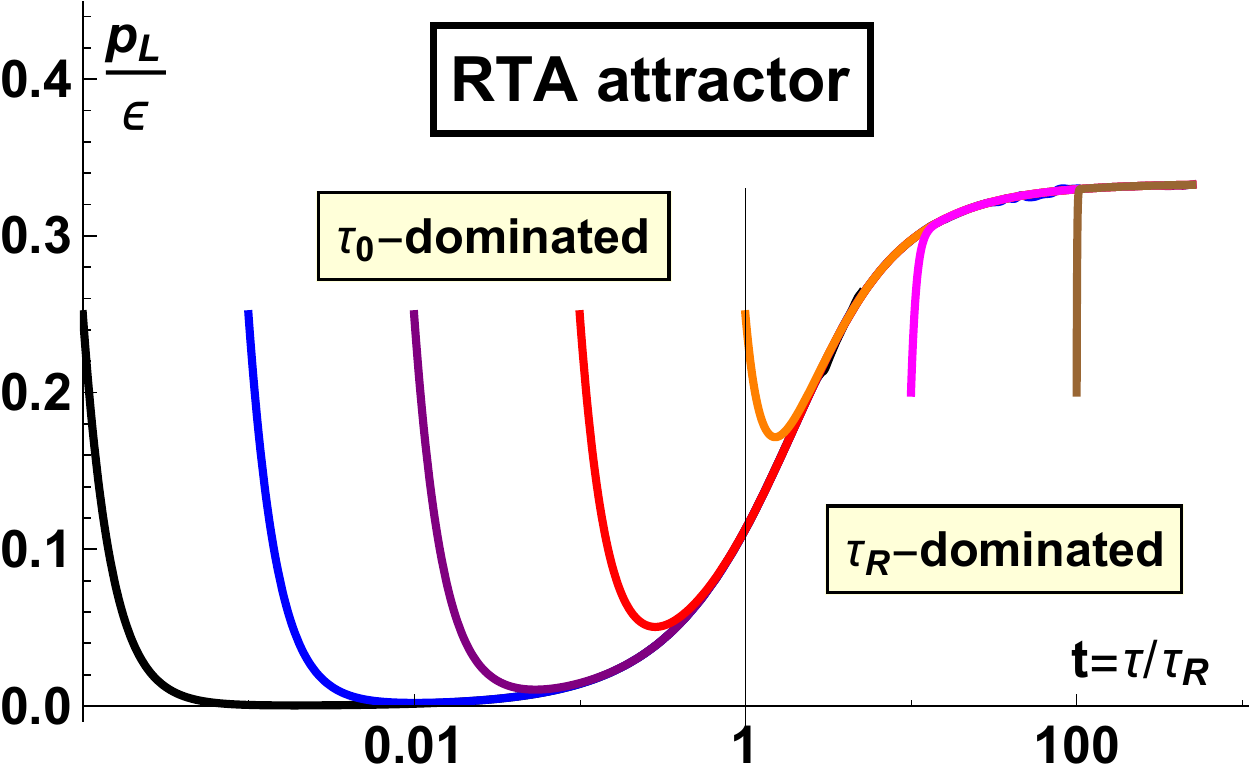}
\caption{The RTA attractor. The ratio of the longitudinal pressure to energy density is plotted for various initializations of \eqref{moment-att}.  Figure from \cite{Kurkela:2019set}.}
\label{fig:kt}
\end{figure}

Depending on the initialization time, the approach to the attractor, which we call $x_A,$ has two qualitatively different regimes, as can be seen in Fig.~\ref{fig:kt}. When the initial time is chosen to be $t_0 \gg 1$, the attractor is approached exponentially quickly, while for $t_0\ll 1$ the approach to the attractor follows approximately a power law decay, $x-x_A\sim t^{-8/3}$. \com{Similar behavior, namely a transition from power law to exponential behavior, is also seen in the case of when an approximate analytic attractor is known \cite{Jaiswal:2019cju}.}

It is worth bearing in mind that the number of moments is infinite. It is beneficial to understand how the attractor might appear at higher moments in the distribution function. Work in the Bjorken flow in the RTA \cite{Strickland:2018ayk,Strickland:2019hff} indicates that the higher moments all approach the attractor at late times, although higher moments {with $l=0$ in} \eqref{moments} approach the exact attractor solution much slower than the other modes.

\subsection{Exact solution in the RTA}

Another method to study the attractor in kinetic theory is the use of the exact solution of the Boltzmann RTA, following e.g.~\cite{Strickland:2018ayk}. Note that an analytic solution in terms of confluent hypergeometric functions was found in the RTA with massless particles with a relaxation time related to the proper time via $\tau_R\propto \tau^{1-\Delta}$, where $\Delta$ is a constant \cite{Blaizot:2020gql,Blaizot:2021cdv}. As in the hydrodynamic case, we will limit our initial consideration to $0+1$ dimensional Bjorken expansion. The RTA attractor has been further examined in \cite{Strickland:2019hff}.

We begin by taking the flow velocity to be
\begin{align}
z^\mu=(z \tau^{-1},0,0,t\tau^{-1}),
\end{align}
which is clearly spacelike $z^\mu \eta_\mn z^\nu=({t^2-z^2})/{\tau^2}=1$.
Due to our assumption of boost invariance, the distribution function will depend only on the proper time, $\tau$, the transverse momenta, $\vec{p}_T$, and the boost-invariant variable  \cite{PhysRevD.30.2371,Bialas:1987en}
\begin{align}
\omega=t p_L- zE,
\end{align}
where $E=p^0.$
One can then define the quantity
\begin{align}
v(\tau,\omega, \vec{p}_T)=E t -p_L z=\sqrt{\omega^2+(m^2+\vec{p}_T^2)\tau^2},
\end{align}
We will consider massless particles, setting $m=0$ in what follows. The above relations lead to a straightforward identification of the energy and longitudinal momentum of a particle as
\begin{align}
E \tau^2&=vt +\omega z,\\
p_L \tau^2 &=\omega t+ v z.
\end{align}
Note that in this case, the integration measure reads
\begin{align}
dP&=\frac{d^4 p}{(2\pi)^4}2 \pi \delta(p^2-m^2)2 \theta(E),\\
&=\frac{d p_L d^2 p_T}{(2\pi)^3 E}=\frac{d\omega d^2 p_T}{(2\pi)^3v}
\end{align}

With the simplifying assumptions and parameterizations outlined above, the RTA Boltzmann equation takes the simple form
\begin{align}
 \partial_\tau  f=\frac{f_{(eq)}-f}{\tau_R(\tau)},\label{RTA}
\end{align}
where we note that the relaxation time $\tau_R$ is a function of proper time.
The RTA Boltzmann equation has a formal solution \cite{BAYM198418,Florkowski:2013lza} 
given by 
\begin{align}\label{exactf}
f(\tau,\omega,p_T)= D(\tau,\tau_0) f_0(\omega,p_T)
+\int^\tau_{\tau_0}\frac{\text{d}\tau^\prime}{\tau_{\rm eq}(\tau^\prime)}D(\tau,\tau^\prime)f_{\rm eq}(\tau^\prime,\omega,p_T),
\end{align}
where the damping function is given by
\begin{align}\label{damp}
D(a,b)=\exp\left[-\int^a_b \frac{\text{d}\tau^\prime}{\tau_{\rm eq}(\tau^\prime)}\right]
\end{align}
and $f_0$ is an arbitrary initial distribution function. Note that the first term in \eqref{exactf} carries much of the information of the initial conditions at early times, but at times greater than $\tau_R$, this first term decays exponentially. Thus the second term describes the late-time behavior of the system.
Finally, note that for conformal RTA, we have \cite{Denicol:2010xn,Denicol:2011fa}
\begin{align}
\tau_{\rm eq}=\frac{5{\eta}}{sT(\tau)}.
\end{align}

To introduce anisotropy, one can use the Romatschke-Strickland distribution function \cite{Romatschke:2003ms} with anisotropy parameter, $\xi$,
\begin{align}\label{aniso-dist}
f=f_{eq}\left(\sqrt{p^2_\perp+(1+\xi) p_z^2}\right),
\end{align}
where without loss of generality, we aligned the anisotropy in the $z$ direction. This was studied in the RTA and with a $\lambda \phi^4$ scalar collision kernel \cite{Almaalol:2018ynx}, which was extended further by considering dynamical fugacity in the RTA \cite{Strickland:2019hff}. Interestingly, models with number-conserving kernels were found to approach the attractor more rapidly, especially the lowest lying mode, although that was not the case for the aHydro attractor \cite{Almaalol:2018jmz}.

One proceeds to solve the equations with the integral iterative method, first outlined in this context in \cite{Florkowski:2013lza,Florkowski:2013lya}. It essentially boils down to integrating both sides of \eqref{exactf} over the momentum squared to arrive at an expression for the energy density. Using the matching conditions, one knows that $\varepsilon=\varepsilon_{\rm eq}$. Then one can read off the temperature from $\varepsilon\sim T^4$. With the temperature, the equilibrium distribution function is known at all times and so the integration can be done in \eqref{exactf} to find the distribution function, $f$.

\subsection{Adiabatic hydrodynamization}

Another interesting approach was to study a system with a gapped Hamiltonian, where the RTA Boltzmann equation of the momentum-weighted distribution function (integrating out the modulus $p=\sqrt{p^2_T+p^2_z}$, but not performing the angular integration) is recast as a time dependent Schr{\"o}dinger equation \cite{Brewer:2019oha}. The lowest eigenmode 
is identified as the pre-hydrodynamic mode, which eventually evolves into the hydrodynamic mode, remaining gapped from the faster modes throughout the evolution. In this set-up, the approach to the attractor is mainly characterized by such pre-hydrodynamic modes, which the authors of \cite{Brewer:2019oha} refer to as adiabatic hydrodynamization. This is due to adiabaticity, i.e. the excitations of the pre-hydrodynamic modes are suppressed. Future work involves relaxing the assumption of the RTA to a more realistic case, e.g. a Fokker-Planck collision kernel. To make better contact with the QGP, it is useful to go beyond the Bjorken expansion by considering transverse modes. A possible experimental signature could be found in the flow coefficients $v_n$ of small systems, where the pre-hydrodynamic modes might not have the chance to fully hydrodynamize.

\subsection{Hydrodynamic generators}

An intriguing notion that has been raised is the existence of hydrodynamic generators \cite{McNelis:2020jrn,McNelis:2021zsp}, where the exact solution of the RTA Boltzmann equation at late times generates the Borel-resummed Chapman-Enskog expansion.
The Chapman-Enskog expansion is an expansion in the Knudsen number, $K_N$, which is a dimensionless parameter given by the ratio of the mean free path and a characteristic macroscopic length scale, particular to the system at hand. For the RTA Boltzmann equation \eqref{RTA}, the Chapman-Enskog expansion is \cite{McNelis:2020jrn}
\begin{align}
f_{CE}(\tau,p)=\sum_{n=0}^\infty (-\tau_R(\tau)\partial_\tau)^n f_{eq}(\tau,p).
\end{align}
This series was found to be factorially divergent \cite{PhysRevLett.56.1571,Denicol:2016bjh} even for small Knudsen number. One can try to Borel resum the Chapman-Enskog series as outlined in Sec.~\ref{sec:borel} to find
\begin{align}\label{borelCE}
f_{CE}(\tau,p)=\int_0^\infty dz e^{-z}\sum_{n=0}^\infty \frac{z^n}{n!}(-\tau_R(\tau)\partial_\tau)^n f_{eq}(\tau,p).
\end{align}
Using the exact solution  \eqref{exactf} in the case the relaxation time is kept constant, after some algebra one finds
\begin{align}\label{midstep}
f(\tau,p)=e^{-z_0}f_0(\tau_0,p)+\int_0^{z_0}d z  e^{-z}\sum_{n=0}^\infty  \frac{(-z \tau_R)^n}{n!} \partial_\tau^n f_{eq}(\tau,p),
\end{align}
where $z_0=({\tau-\tau_0})/{\tau_R}.$ Clearly, expanding \eqref{midstep} for large proper time reduces to \eqref{borelCE}.
From here, the authors  conjecture that even in the case the relaxation time is proper time dependent, the hydrodynamic generator is given by \cite{McNelis:2020jrn} 
\begin{align}
f_G(\tau,p)=\int_{\tau_0}^\tau \frac{d\tau^\prime D(\tau,\tau^\prime)f_{\rm eq}(\tau^\prime,p)}{\tau_r(\tau^\prime)},
\end{align}
where $D(\tau,\tau^\prime)$ is the damping factor \eqref{damp}. The hydrodynamic generator is a nice representation of the Chapman-Enskog expansion as it is finite, is amenable to generalizations to $(3+1)$ dimensions and to systems without Bjorken flow, and satisfies the RTA Boltzmann equation
\begin{align}
\partial_\tau f_G=\frac{f_{\rm eq}-f}{\tau_R(\tau)}.
\end{align}
Interestingly, the RTA hydrodynamic generator contains the Chapman-Enskog expansion with coupled non-hydrodynamic modes with different decay times. It is found that the higher order gradient corrections are suppressed during hydrodynamization, which indicates that these higher corrections do not necessarily play a large role, even with large gradients. This story seems to resemble the modes found via resurgence in \cite{Heller:2016rtz,Heller:2018qvh}.

Finally, it should be pointed out that the RTA has its limitations. The derivation above rested on the knowledge of an exact solution in the RTA, so incorporating more realistic collision kernels will require the damping function \eqref{damp} to be replaced more generally with a Green's function, see e.g. \cite{Kamata:2020mka}.

\section{An eye to phenomenology}\label{sec:pheno}

We have seen that hydrodynamic attractors are found in many systems. The study of these attractors offer plenty of theoretical insights, e.g. to the applicability of hydrodynamics, resurgence and transseries, but what about about real world applications? In this section, we highlight some of the phenomenology in heavy ion physics that attractors played an important part.

\subsection{K{\o}MP{\o}ST}

\begin{figure}\centering
\includegraphics[width=0.75\linewidth]{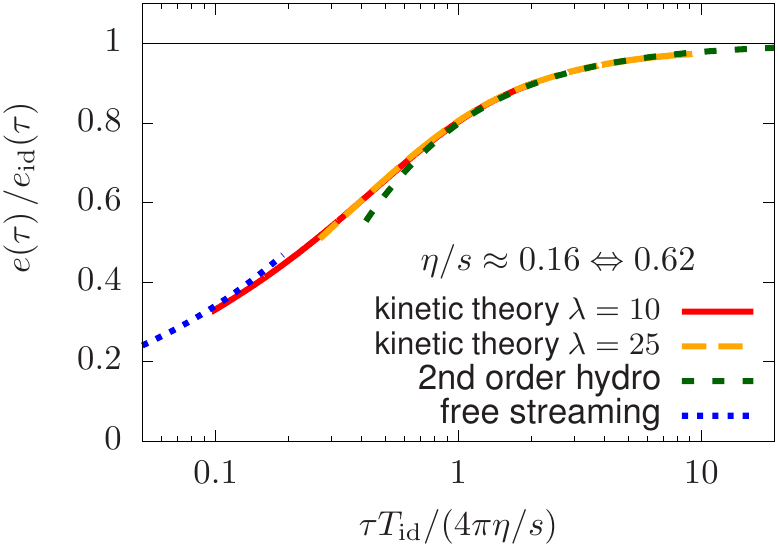}
\caption{Evolution of the background energy density divided by the ideal hydrodynamic energy density \eqref{teff} as a function of proper time (normalized by the kinetic relaxation time $\tau_R\equiv (\eta/s )/ T_{\rm id})$.  Note that attractor curve goes smoothly from the early time free streaming limit to viscous hydrodynamics at late times. Figure from \cite{Kurkela:2018wud}.}
\label{fig:kompost}
\end{figure}

K{\o}MP{\o}ST is a framework that matches the early time, nonequilibrium QCD kinetic theory to a later time when a hydrodynamic description is more appropriate \cite{Kurkela:2018wud,Kurkela:2018vqr}. The initial energy momentum tensor is described by kinetic theory beginning at the time scale, $\tau_{\rm EKT}$. This is then propagated to later times when hydrodynamics is applicable, $\tau_{\rm hydro}$, via nonequilibrium Green's functions, $G^\mn_\ab$, which are computed in QCD kinetic theory. The linearized perturbations of the stress tensor
\begin{align}
T^\mn(\tau_{\rm EKT},\textbf{x})=\bar T^\mn (\tau_{\rm EKT})+\delta T^\mn(\tau_{\rm EKT},\textbf{x}),
\end{align}
evolve according to 
\begin{align}
\delta T^\mn(\tau_{\rm hydro},\textbf{x})=\int d^2 x^\prime G^\mn_\ab (\textbf{x}, \textbf{x}^\prime, \tau_{\rm hydro},\tau_{\rm EKT}) \delta T^\ab(\tau_{\rm EKT},x^\prime)\frac{\bar T^{\tau\tau}(\tau_{\rm hydro})}{\bar T^{\tau\tau}(\tau_{\rm EKT})}.
\end{align}
The barred quantities denote the background energy momentum tensor, which is computed by taking a spatial average over the causal circle.
As observed in 
\cite{Keegan:2015avk,Heller:2016rtz,Kurkela:2018wud}, the dynamics in these timescales becomes essentially independent of the microscopic kinetic coupling when measured in (macroscopic) dimensionless time
\begin{align}\label{omegabar}
\bar{\omega}=\frac{\tau T}{4\pi \eta/s}.
\end{align}
Even far from equilibrium for $\bar{\omega}<1$, it is observed that the energy momentum tensor is a function of $\bar{\omega}$ only with hydrodynamization occuring after $\bar{\omega}>1.$ As such, the attractor can be viewed as a bridge between the kinetic and hydrodynamic regimes. The universal attractor behavior can be seen in Fig.~\ref{fig:kompost}, where two values of the kinetic theory coupling is plotted for a range of $\eta/s$. The nonequilibrium evolution of the system goes from early time free streaming to late time viscous hydrodynamics. The universal scaling function is due to the attractor behavior.

K{\o}MP{\o}ST has been incorporated in a number of contexts, e.g. to pre-hydrodynamic evolution and its signatures in final-state heavy-ion observables \cite{NunesdaSilva:2020bfs} and more recently, to describe photon observables in small collision systems \cite{Gale:2021zlc}.

\subsection{Entropy production}

The fact that hydrodynamic attractors can serve as a bridge between late-time and earlier, far-from-equilibrium behavior was recently exploited to extract phenomenologically relevant information in the QGP context \cite{Giacalone:2019ldn}. The central idea is that the multiplicity of final state particles measured in detectors is intimately linked to the entropy production during the pre-equilibrium phase. The total entropy per unit spacetime rapidity, $\eta_s,$ during the near-equilbrium transverse expansion of the QGP is 
\begin{align}
\frac{dS}{d \eta_s}=A_\perp (s\tau)_{\rm hydro},
\end{align}
which is related to the charged particle multiplicity at late times via
\begin{align}\label{mult}
\frac{dN_{\rm ch}}{d \eta_s} \frac{d S}{d y}\approx \frac{1}{J}A_\perp (s\tau)_{\rm hydro} \frac{dN_{\rm ch}}{dy},
\end{align}
where $J$ denotes the Jacobian factor, $y$ is the rapidity, \textcolor{black}{$dy/d\eta\approx 1.1$}, $A_\perp=\pi R^2$ and $2R$ is the transverse extent of the system.
Then the initial energy density can be linked to the experimentally observed particle multiplicities via the neat pocket formula \cite{Giacalone:2019ldn}
\begin{align}\label{pocket}
(s\tau)_{\rm hydro}= \frac{4}{3}C_\infty^{3/4}\left(4\pi \frac{\eta}{s}\right)^{1/3} \left(\frac{\pi^2}{30}\nu_{\rm eff}\right)^{1/3}(\varepsilon \tau)_0^{2/3},
\end{align}
where $C_\infty$ is a constant of order unity and $\nu_{\rm eff}$ is the number of effective degrees of freedom. For an ideal gas of gluons, $\nu_{\rm eff} =16$, while for a mixture of quarks and gluons, $\nu_{\rm eff} =47.5$.

To derive this result, the knowledge that there is an attractor can be put to work to find 
\begin{align}\label{e-attr}
\frac{\varepsilon(\tau)\tau^{4/3}}{\varepsilon_{\rm hydro}\tau_{\rm hydro}^{4/3}}=\mathcal{E}(\bar \omega),
\end{align}
where the quantity on the right hand side is the attractor curve. 
To see this, we follow the derivation of the link between the initial state energy density, $\varepsilon_0$, and the near thermal system energy density, $\varepsilon_{\rm hydro}$, beginning from the boost-invariant conservation law
\begin{align}\label{bjork-consv}
\partial_\tau \varepsilon =- \frac{\varepsilon+P_L}{\tau}.
\end{align}
The energy density is related to the effective temperature via 
\begin{align}\label{teff}
\varepsilon(\tau)=\frac{\pi^2}{30}\nu_{\rm eff} T^4_{\rm eff}(\tau),
\end{align} 
It is then convenient to introduce the dimensionless scaling variable $\bar{\omega}$ given by \eqref{omegabar}.
Finally, calling the anisotropy $f(\bar{\omega})\equiv P_L/\varepsilon$, we see that we can rewrite \eqref{bjork-consv} as
\begin{align}
\frac{d\bar{\omega}}{d\tau}\partial_{\bar{\omega}} \log \varepsilon=-\frac{1+f}{\tau}.
\end{align}
Using the definition of $\bar{\omega}$ and the relation between the energy density and the effective temperature \eqref{teff},
we see that 
\begin{align}
\frac{4\pi \eta}{s}\frac{d\bar{\omega}}{d\tau}&=T_{\rm eff}+\tau \partial_\tau T_{\rm eff},\\
&=T_{\rm eff}+\frac{1}{4} T_{\rm eff}(\tau \varepsilon^{-1} \partial_\tau  \varepsilon),\\
&=T_{\rm eff}\left(\frac{3}{4}-\frac{1}{4}f \right).
\end{align}
Thus, 
\begin{align}
\frac{\partial}{\partial \bar{\omega}} \log \varepsilon
=-\frac{1+f}{\frac{\tau T_{\rm eff}}{4\pi\eta/s}\left(\frac{3}{4}-\frac{1}{4}f\right)}
=-\frac{1+f}{\bar{\omega}\left(\frac{3}{4}-\frac{1}{4}f\right)}.
\end{align}
After integration, we arrive at 
\begin{align}
\varepsilon(\tau)=\varepsilon_0 \exp\left( -\int_{\bar{\omega_0}}^{\bar{\omega}_{\rm hydro}} \frac{d\bar{\omega}}{\bar{\omega}}\frac{1+f}{\frac{3}{4}-\frac{1}{4}f}\right).
\end{align}
From here, we arrive straightforwardly at \eqref{e-attr}. Note that the energy attractor \eqref{e-attr} interpolates smoothly between free streaming and late time viscous hydrodynamics
\begin{align}
\mathcal{E}(\bar\omega\ll1)=C_\infty^{-1}\bar\omega^{4/9},\\
\mathcal{E}(\bar\omega\gg1)=1-\frac{2}{3\pi\bar\omega},
\end{align}
respectively. 
The pocket formula \eqref{pocket} and the charged particle multiplicity \eqref{mult} can be used to estimate the initial energy density for central lead collisions at $\sqrt{s_{NN}}=2.76$ TeV to be $\varepsilon\approx 270$ GeV/fm$^3$ at the time $\tau_0=0.1$ fm/c \cite{Giacalone:2019ldn}.
Furthermore, the study of such energy attractors has been used to characterize potential probes of the QGP at early times, such as intermediate mass dileptons \cite{Coquet:2021cuv}. \com{Interestingly, the evolution of a system following an attractor solution has been shown to result in higher yields (in fact, maximum yields) in thermal dileption and photon particle production \cite{Naik:2021yph}, compared to solutions away from the attractor. }

The previous argument was turned on its head in \cite{Jankowski:2020itt}, where the assumption of free streaming at early times was explored by positing the energy density scales with some to-be-determined parameter, $\beta$, namely $\varepsilon\sim \tau^{-\beta}$ (the free streaming case has $\beta=1$). Using different initial state models and comparing the final state to experimental data, they find that the preferred expansion parameter $\beta$ is slightly, but distinctly different from the free streaming case.

\subsection{Non-zero chemical potential}\label{sec:chem-potential}

Real world QGP has non-zero baryon chemical potential. As such, it is natural to include the effect of non-zero chemical potential on the attractor story described so far. This has been explored both in hydrodynamic \cite{Dore:2020fiq,Dore:2020jye} and kinetic setups \cite{Du:2020zqg,Du:2020dvp}. In the hydrodynamic case, the $0+1d$ DNMR Bjorken case described in Sec.~\ref{bjork-example} is supplemented with an evolution of the baryon number density and relaxation of the bulk viscosity. The late-time attractor regime was not reached due to the short lifetime of the hydrodynamic runs, so it is more apt to talk of potential attractors, which the system looked like it was approaching. Although attractors were not systematically studied, there was the indication that near the critical point, the bulk attractor should have a pronounced effect. Also, the potential attractor line has a different behavior when comparing MIS to DNMR near the critical point, indicating care should be taken as there are phenomenological implications and differing results depending on which set of equations of motion is used \cite{Dore:2020jye}. 

In the kinetic approach of \cite{Du:2020zqg,Du:2020dvp}, quark/anti-quark and gluonic distribution functions are taken into account, 
which respectively take the form
\begin{align}
f_{q_f/\bar{q}_f}&=f_{q_f/\bar{q}_f}^0\frac{\sqrt{p_T^2+p_\parallel^2}}{\sqrt{p_T^2+\xi_0^2 p_\parallel^2}}e^{-2(p_T^2+\xi_0^2p_\parallel^2)/(3Q_0^2)},\\
f_g&=\frac{{f_g^0 Q_0}}{\sqrt{p_T^2+\xi_0^2 p_\parallel^2}}e^{-2(p_T^2+\xi_0^2p_\parallel^2)/(3Q_0^2)},
\end{align}
where $\xi_0$ is the initial anisotropy and $Q_0$ is the initial energy scale.
The evolution of the anisotropic distribution functions with form similar to \eqref{aniso-dist} is given by the QCD Boltzmann equation \cite{Mueller:1999pi} with a collision kernel with leading order elastic ($2\leftrightarrow2$) and inelastic ($1\leftrightarrow2$) interactions taken into account, i.e. for a distribution function of species $a$, $f_a=f_a(\tau,p_T,p_\parallel)$
\begin{align}
\left[\partial_\tau- \frac{p_\parallel}{\tau}\frac{\partial}{\partial p_\parallel} \right]f_a=-C^{2\leftrightarrow2}[f]-C^{2\leftrightarrow1}[f].
\end{align}
The evolution of the system is given by the conservation of the energy momentum tensor and the conserved current, $\Delta J^\mu_f=(\Delta n_f,0,0,0)$, via
\begin{align}
0&=\partial_\tau \varepsilon+\frac{\varepsilon+p_L}{\tau},\\
0&=\partial_\tau \Delta n_f+\frac{\Delta n_f}{\tau}.
\end{align}
The nonequilibrium system has a build up of longitudinal pressure due to interactions while the longitudinal expansion slows down, before finally at late times approaching local equilibrium with $p_L=p_T=\varepsilon/3$.

Like in the case of zero chemical potential discussed previously, the validity of viscous hydrodynamics as a description of the non-equilibrium evolution in the case of non-zero chemical potential also begins at timescales roughly given by $ 4\pi \eta/(sT) $   \cite{Du:2020zqg,Du:2020dvp}. However, at early times non-zero chemical potential can have a large impact on the pressure for times earlier than $ 4\pi \eta/(sT) $, although the late time approach is found to be affected by around $~10\%$. This is to be contrasted with results from kinetic theory computations involving a single component distribution function, as discussed in Sec.~\ref{sec:kt}, where the rapid memory loss of the initial conditions characteristic to attractor behavior allows for a universal approach to viscous hydrodynamics. As such, it is clear that the chemical composition of the QGP has an effect on the approach to the universal approach to viscous hydrodynamics at late time.

\subsection{Improved freeze-out procedures}

 It is worthwhile to keep in mind that for all the discussion of hydrodynamic variables as a means to describe heavy ion collisions, we have left out a crucial ingredient. Experiments themselves measure the distribution of particles and not the fluid-dynamic variables directly as discussed previously. Following the collision and initial evolution of the quark gluon plasma, the distribution of particles ``freeze-out'' as color neutral hadrons and free stream towards the detectors at time scales around $\tau\sim 10 \text{ fm/c}$. The translation of distribution of particles to hydrodynamic variables is known as the freeze-out procedure.

Recently, it was pointed out in \cite{Almaalol:2020rnu} that possible improvements to freeze-out procedures can be made in light of the developments surrounding the attractor. Using Arnold-Moore-Yaffe (AMY) effective kinetic theory in one-dimensional Bjorken flow, where the collision kernel includes gluon scattering and inelastic gluon splitting using the pure glue code from \cite{PhysRevLett.115.182301}, a comparison was made between aHydro and linearized freeze-out prescriptions. It was found that the freeze-out prescriptions studied tend to reproduce the low-order moments of the distribution at late times $\tau\gg 5\text{ fm/c}.$ However, for higher moments, it was found that the linearized prescriptions do poorly compared to aHydro. The last fact indicates that the aHydro freeze-out prescription can be of use for the study of smaller systems, like $pA$ or peripheral $AA$ collisions.

\subsection{Beyond conformal symmetry}

The majority of the work discussed so far centered on conformal systems, which provides the advantage of theoretical clarity and sometimes the feasibility of analytical computations. However, in more realistic systems, one should not expect full conformal symmetry to hold, especially with a full three dimensional expansion and a realistic equation of state. There are a number of works that go beyond the status of the field, including the exploratory work of attractors with a non-conformal equation of state relevant for the QCD critical point considered in \cite{Dore:2020fiq,Dore:2020jye}, see Sec.~\ref{sec:chem-potential} for more details. 

Here we discuss an initial survey that was made in \cite{Romatschke:2017acs} for a $0+1$-dimensional 
Bjorken kinetic theory of a gas of particles of mass $m$ and for inhomogeneous $2+1$ dimensional BRSSS hydrodynamics. Two quantities were proposed to investigate whether attractor solutions are present, namely
\begin{align}\label{a1eq}
A_1&=\frac{u^\mu \nabla_\mu \varepsilon}{(\varepsilon+P)\nabla_\lambda u^\lambda},\\
A_2&=\frac{\text{Tr} T^\mn+\varepsilon-3P}{\zeta T}.
\end{align} 
The first quantity represents a simple rewriting of the evolution of the energy density from the Navier-Stokes equations, while the second denotes the deviation of the equation of state from the conformal limit. When expressed as a function of the inverse gradient strength $\Gamma$,
\begin{align}
\Gamma=\left[\frac{\eta}{2s}\frac{\sigma^\mn \sigma_\mn}{T\nabla_\lambda u^\lambda}+\frac{\zeta}{s}\frac{\nabla_\lambda u^\lambda}{T}\right]^{-1},
\end{align}
both $A_1$ and $A_2$ exhibit attractor behavior. However, the approach of different solutions is slower for $A_2$ than $A_1$, which may indicate that at early times, the attractor is harder to discern in non-conformal systems. In the case of the $2+1$ dimensional BRSSS hydrodynamics, the apparent attractor was seen by studying $A_1(\Gamma)$, which seemed to greatly depart from the expected Navier-Stokes limit, although such a simulation proved too computationally costly to extend for much later times.

In many respects, although the exploration of the non-conformal limit is just at beginning, there are a number of lessons that have been learned already. As mentioned above, moving away from the conformal limit also can provide insight into the nature of the early time attractor. Recent work involved studying a $0+1$ dimensional gas of massive particles evolving under the Boltzmann equation in the RTA, which was compared to non-conformal hydrodynamics  \cite{Chattopadhyay:2021ive,Jaiswal:2021uvv}
. It is found that only the longitudinal pressure has an early time attractor, while that there is no early time hydrodynamic attractor for the bulk and shear viscous stresses, adding evidence to the related indication in \cite{Kurkela:2019set}. Moreover, the comparison of the quantity \eqref{a1eq} in \cite{Jaiswal:2021uvv} indicates that while for late times $A_1$ does exhibit universal attractor behavior, at early times there is a noticeable difference between the solutions depending on the choice of different masses. 
Another interesting observation in non-conformal Bjorken flow is that it is not a guarentee one even reaches local thermal equilibrium at late times. For instance, in the case of variable relaxation time scaling, namely $\tau_{R}\propto T^{-\Delta}$ ($\Delta=1$ corresponds to the conformal limit), the system can evolve to a late time regime out of local thermal equilibrium for $\Delta\neq1$ \cite{Heller:2018qvh}, in contrast to conformal Bjorken flow.

\section{Conclusion and Outlook}

The story of hydrodynamic attractors in heavy ion collisions is still being written. We can take this point to reflect on where the field stands and where it might go.  At this stage, it seems like attractors for Bjorken flow are well-established for a variety of effective theories relevant for heavy ion collisions, such as relativistic hydrodynamics, holography and kinetic theory. Attractors provide a conceptual mechanism to understand the applicability of hydrodynamics even in far-from-equilibrium systems, as well as providing some phenomenological insight. 
Notwithstanding the current successes, there is still much work to be done. Some open questions remain:
\begin{itemize} 
\item  \textbf{What are the theoretical properties of the QCD attractor and under which conditions does QCD exhibit
attractor behaviour?}
\item \textbf{How do attractors behave in systems with relaxed symmetries?} Studies of attractors have begun to go beyond the Bjorken and Gubser cases described here.
Ultimately, the goal should be to consider the full symmetries in heavy ion collisions, moving away from the conformal limit as in \cite{Romatschke:2017acs,Chattopadhyay:2021ive,Jaiswal:2021uvv,Chen:2021wwh} and including transverse expansion in $3+1$ dimensions, as well as the quark degrees of freedom as in \cite{Kurkela:2018xxd,Kurkela:2018oqw,Dore:2020jye,Du:2020zqg,Du:2020dvp}. 
There is some promising work to take transverse dynamics into account \cite{Ambrus:2021sjg}. Recently, the attractor was studied in a Hubble expanding background \cite{Du:2021fok}, where the scale factor was assumed to follow a power law form $a(t)\sim t^\alpha$ instead of solving the Einstein equations to find $a(t)$, an assumption informed by cosmology.
\item \textbf{Beyond numerical techniques, how can we extend the determination of attractors to earlier times?} Especially in the more complicated setting of relaxed symmetries, it is not clear if there is an unambiguous early time attractor.  Also, in the holographic context, it seems like there is only the late-time attractor \cite{Kurkela:2019set} (although the next question also can be relevant for the early time attractor).
\item \textbf{How would finite coupling corrections affect the story of holographic attractors?}  The Gauss-Bonnet setup described in Sec.~\ref{sec:GB} can act as a laboratory to study higher derivative corrections. The observation that varying a single parameter allows the interpolation from the strongly coupled to the expectation from weakly coupled kinetic theory should be viewed as an encouraging step in this direction. It can be speculated that finite coupling corrections thus might provide an unambiguous method of determining the possible early time behavior of the attractor. 
\item \textbf{How do hydro attractors fit in with non-thermal fixed points?} It would be interesting to see a study of the evolution of a kinetic model from a non-thermal fixed point to a late time hydrodynamic attractor.
\item \textbf{Are there any signatures of attractors in smaller systems, e.g. $pp$ collisions?} In $pp$ collisions, the number of particles produced is regularly quite low compared to typical heavy ion collisions. It is not clear if hydrodynamization would occur in such short-lived settings \cite{Wiedemann:2021wju}. If so, there could still be the approach to the hydrodynamic attractor for generic initial conditions without reaching the attractor at all, potentially indicating some level of pre-hydrodynamization in the smaller systems.   
\item \textbf{Which observables are sensitive to hydrodynamic attractors?} Attractors seem to be a useful tool in using final state information to probe earlier times in heavy ion collisions, see Sec.~\ref{sec:pheno}. Furthermore, one can envisage using them to determine which final state observables depend directly on the initial state. 

\item \textbf{How universal are hydrodynamic attractors?} The hydrodynamic limit exists in a number of neighboring fields, including cosmology and condensed matter systems. The lessons learned about attractors in the high energy community can surely be applied to expanding systems in a variety of fields.
\end{itemize}

\subsection*{Acknowledgements}
I would like to extend my gratitude to Kirill Boguslavski, Eduardo Grossi, Michal Heller, Aleksi Kurkela, Mauricio Martinez, Aleksas Mazeliauskas, Ayan Mukhopadhyay, David M{\"u}ller, Jorge Noronha, 
Anton Rebhan, Paul Romatschke,  Alexandre Serantes, 
Soeren Schlichtling,
 Micha\l~Spali{\'n}ski, Michael Strickland, Victor Svensson, Derek Teaney, Urs Wiedemann and Yi Yin
for helpful discussions.
Many thanks to Eduardo Grossi, Ayan Mukhopadhyay and David M{\"u}ller
 for comments on the manuscript.
This work is supported by the Erwin Schr{\"o}dinger Fellowship of the Austrian Science Fund (FWF), project no. J4406.

\appendix

\section{List of abbreviations}\label{app:abbrev}

\begin{description}[labelsep=6cm, align=left]
\item [3+1D]		3+1 dimensions / dimensional
\item [AdS/CFT]		Anti-de Sitter/Conformal Field Theory
\item [AMY]		Arnold-Moore-Yaffe
\item[BGK]			Bhatnagar-Gross-Krook
\item[BRSSS]		Baier-Romatschke-Son-Starinets-Stephanov
\item[DNMR]		Denicol-Niemi-Molnar-Rischke
\item[GB] 			Gauss-Bonnet
\item [HJSW]		Heller-Janik-Spalinski-Witaszczyk 
\item [K{\o}MP{\o}ST] Kurkela-Mazeliauskas-Paquet-Schlichtling-Teaney
\item [LHC]		Large Hadron Collider
\item [MIS]		M{\"u}ller-Israel-Stewart
\item [QCD]		quantum chromodynamics
\item [QGP]		quark-gluon plasma
\item [RHIC]		Relativistic Heavy Ion Collider
\item [RTA]		relaxed time approximation
\item [SYM]		Super	Yang-Mills
\end{description}

\bibliography{tutorial}
\bibliographystyle{JHEP}

\end{document}